\documentclass[letterpaper, 10 pt, journal, twoside]{IEEEtran}

\usepackage[utf8]{inputenc}
\usepackage{blindtext}
\usepackage{tabularx}

\usepackage{amsthm}

\usepackage{mathtools}
\usepackage{amssymb}
\usepackage{cite}
\usepackage{xcolor}
\usepackage{tikz}
\usepackage{pgfplots}
\usepackage{svg}
\usepackage{adjustbox}
\usepackage{comment}
\usepackage{standalone}
\usepackage{calc}
\usepackage[acronym]{glossaries}
\usepackage{ifthen}
\newboolean{proofs}

\usepackage[
 colorlinks,
    linkcolor={blue!50!black},
    citecolor={blue!50!black},
    urlcolor={blue!80!black}]{hyperref}
    
\usepackage{multicol}
\usepackage{todonotes}
\presetkeys{todonotes}{inline}{}

\usepackage[capitalize]{cleveref}
\Crefname{equation}{Eq.}{Eqs.}
\Crefname{figure}{Fig.}{Figs.}
\Crefname{tabular}{Tab.}{Tabs.}
\Crefname{definition}{Def.}{Defs.}
\Crefname{section}{Sec.}{Sects.}
\Crefname{theorem}{Thm.}{Thms.}
\Crefname{condition}{Cond.}{Conds.}

\usepackage{graphicx}
\usepackage{booktabs}
\usepackage{threeparttable}
\usepackage{multirow}
\usepackage[ruled,vlined]{algorithm2e}
\usepackage{tikz, pgfplots, pgfplotstable}
\usetikzlibrary{
    positioning,
  	cd,
  	shapes,
  	math,
  	decorations.markings,
  	decorations.pathmorphing,
	decorations.pathreplacing,
  	arrows.meta,
  	calc,
  	fit,
  	quotes,
  }
\tikzcdset{arrow style=tikz, diagrams={>=To}}
\pgfplotsset{compat=1.15}

\tikzstyle{block} = [draw, rectangle, minimum height=2em, minimum width=3em,thick]

\tikzstyle{blockdot} = [block, dotted,rounded corners=4, inner sep=-2pt]

\tikzstyle{blockfill} = [block,rounded corners=4, inner sep=-2pt,fill=blue!5!white]

\tikzstyle{every node}=[font=\footnotesize]

\makeatletter
\let\MYcaption\@makecaption
\makeatother
\usepackage[font=footnotesize]{subcaption}
\makeatletter
\let\@makecaption\MYcaption
\makeatother

\theoremstyle{definition}
\newtheorem{theorem}{Theorem}

\newtheorem{condition}[theorem]{Condition}
\newtheorem{lemma}[theorem]{Lemma}

\newtheorem{remark}[theorem]{Remark}
\newtheorem{definition}[theorem]{Definition}

\makeatletter
\def\@opargbegintheorem#1#2#3{\trivlist
   \item[]{\bfseries #1\ #2\ (#3)} \itshape}
\makeatother

\newacronym{ibr}{IBR}{iterated-best-response}
\newacronym{nash}{NE}{Nash Equilibria}
\newacronym{cudg}{CUDG}{Communal Urban Driving Game}
\newacronym{cudgs}{CUDGs}{Communal Urban Driving Games}
\newacronym{gnep}{GNEP}{Generalized Nash Equilibrium Problem}

\newcommand{\argtuple}[2]{#1, #2}

\newcommand{\agents}{\mathcal{A}} %
\newcommand{\aggregation}{\mathrm{agg}}

\newacronym{abk:av}{AV}{Autonomous Vehicle}

\newacronym{abk:br}{BR}{Best Response}
\newcommand{\br}{\mathsf{BR}}
\newcommand{\brstrict}{\br^{\prec}}
\newcommand{\brweak}{\br^{\precsim}}

\definecolor{baiocchi}{RGB}{193,221,245}
\newcommand*\circled[1]{\tikz[baseline=(char.base)]{
            \node[shape=circle,draw=black,inner sep=0.5pt] (char) {#1};}}
\newcommand{\game}{\mathcal{G}}
\newacronym{abk:gpop}{GPOP}{Game with Partially Ordered Preferences}
\newcommand{\height}[1]{\mathsf{height}(#1)}

\newcommand{\coll}{\jointmetric^1}
\newcommand{\driv}{\metric^2}
\newcommand{\mincl}{\jointmetric^3}
\newcommand{\epi}{\metric^4}
\newcommand{\lat}{\metric^5}
\newcommand{\longi}{\metric^6}
\newcommand{\prog}{\metric^7}
\newcommand{\devh}{\metric^8}
\newcommand{\devl}{\metric^9}

\newcommand{\metricsSet}{\mathcal{M}}
\newcommand{\metricsSetComp}{\metricsSet^{\mathrm{c}}}
\newcommand{\allmetrics}{\mathbf{M}}
\newcommand{\metric}{m}
\newcommand{\jointmetric}{\overline{\metric}}
\newcommand{\jointmetricSet}{\overline{\metricsSet}}
\newcommand{\Min}{\mathrm{Min}}

\newcommand{\colltext}{\textsf{collision}}
\newcommand{\drivtext}{\textsf{area violation}}
\newcommand{\mincltext}{\textsf{clearance}}
\newcommand{\longitext}{\textsf{long. comfort}}
\newcommand{\lattext}{\textsf{lat. comfort}}
\newcommand{\devhtext}{\textsf{dev. heading}}
\newcommand{\devltext}{\textsf{dev. lateral}}
\newcommand{\progtext}{\textsf{progress}}
\newcommand{\epitext}{\textsf{time}}

\newacronym{abk:ne}{NE}{Nash Equilibrium}
\newcommand{\nasheq}{\mathsf{NE}}
\newcommand{\neqstrong}{\nasheq^{\prec}}
\newcommand{\neqweak}{\nasheq^{\precsim}}
\newcommand{\neqweakadm}{\nasheq_\mathsf{A}^{\precsim}}

\newcommand{\outcomeset}{O}
\newcommand{\outcome}{o}

\newcommand{\posleq}{\preceq}
\newcommand{\preleq}{\precsim}

\newcommand{\posgeq}{\succeq}
\newcommand{\pre}[1]{\text{pr}(#1)}

\newcommand{\preference}{\mathsf{P}}

\newcommand{\joinofprefs}{\preference^\mathrm{u}}

\newcommand{\players}{\mathcal{A}}
\newcommand{\posreals}{\mathbb{R}_{\geq 0}}

\newcommand{\potential}{\mathsf{Pot}}

\newcommand{\playerone}{\textcolor{red}{\text{P1}}}
\newcommand{\playertwo}{\textcolor{blue}{\text{P2}}}
\newcommand{\playerthree}{\textcolor{black!60!green}{\text{P3}}}

\newcommand{\rankk}{\mathsf{r}}
\newcommand{\rankdec}{\mathsf{R}}
\newcommand{\rankne}{\mathsf{R}}

\newcommand{\tup}[1]{\left\langle#1\right\rangle}

\newcommand{\traj}{\gamma}
\newcommand{\trajSpace}{\Gamma}
\newcommand{\trajset}{\Gamma}
\newcommand{\trajne}{\traj^\star}

\newcommand{\upit}{\uparrow}

\newcommand{\reviewed}[1]{{\color{black}#1}}
\usepackage{ifxetex}
\ifxetex
\usepackage[tracking=false,kerning=false,spacing=false]{microtype}
\usepackage[protrusion=true,tracking=false,kerning=false,spacing=false]{microtype}
\else
\usepackage[tracking=false,kerning=true,spacing=true]{microtype}

\fi
\usepackage{enumitem}
\setlist[enumerate]{itemsep=0mm,leftmargin=4mm,topsep=0mm}
\setlist[itemize]{itemsep=0mm,leftmargin=4mm,topsep=0mm}

\begin{document}
\setboolean{proofs}{true}
\bstctlcite{IEEEexample:BSTcontrol}
\title{\LARGE \bf
	Posetal Games: Efficiency, Existence, and Refinement\\ of Equilibria in Games with Prioritized Metrics
}
\author{Alessandro Zanardi$^{1^\ast}$, Gioele Zardini$^{1^\ast}$, Sirish Srinivasan$^{1}$,\\ Saverio Bolognani$^{2}$, Andrea Censi$^{1}$, Florian D\"{o}rfler$^{2}$, Emilio Frazzoli$^{1}$
\thanks{
$^{*}$The first two authors contributed equally to this work.
This work was supported by the Swiss National Science Foundation under NCCR Automation, grant agreement 51NF40\_180545.}
\thanks{
$^{1}$Institute for Dynamic Systems and Control, ETH Z\"urich, Switzerland {\tt \{azanardi,gzardini\}@ethz.ch}.}
\thanks{
$^{2}$Automatic Control Lab, ETH Z\"urich, Switzerland. %
}}

\maketitle
\begin{abstract}
Modern applications require robots to comply with multiple, often conflicting rules and to interact with the other \reviewed{agents}.
We present Posetal Games as a class of games \reviewed{in which} each player expresses a preference over the outcomes via a partially ordered set of metrics. 
This \reviewed{allows one} to combine hierarchical priorities of each player with the interactive nature of the environment.
By contextualizing standard game theoretical notions, we provide two sufficient conditions on the preference of the players to prove existence of pure Nash Equilibria in finite action sets.
\reviewed{Moreover, we define formal operations on the preference structures and link them to a refinement of the game solutions, showing how the set of equilibria can be systematically shrunk.}
The presented results are showcased in a driving game where autonomous vehicles select from a finite set of trajectories. The results demonstrate the interpretability of results in terms of minimum-rank-violation for each player.
\end{abstract}
\IEEEpeerreviewmaketitle
\section{Introduction}

It is well known that decision making is a stressful task for human beings~\cite{Hughes2017WhenWrong}. 
While robots do not get stressed (yet), their prospective ubiquity in our society is faced with similar challenges, charged with the need to ``make the right choice'' in complex environments. 
Indeed, embodied intelligence has to cope with laws of different severity, unwritten rules, different local cultures, liability issues, and different types of agents.
This poses an unmatched challenge for decision making.

There are two aspects which make the problem difficult.
First, robots' behavior needs to be compliant with rules written by humans for humans~\cite{Censi2019b}.
Such rules are often subject to interpretation and need to be contextualized. 
Second, the designed systems need to be robust to the highly interactive nature of unconstrained environments, for which too conservative approaches simply fail~\cite{Trautman2010}.

A clear example featuring both the aspects is autonomous driving. 
The early prototypes were designed to blindly respect the rules.
Quickly, one realized that this was not enough, as one needed robots to obey the unwritten rules of human interactions to blend in~\cite{schwarting2018planning}. 

\reviewed{Accounting for these aspects singularly comes with strong limitations.}
For instance, it is unclear how a system relying only on learned behaviors and ``common practices'' of human interactions will react when faced with a rare unfortunate situation. 
This creates a serious threat for all the participants involved and for the ego-robot also in terms of liability.
Indeed, the outcome could be catastrophic. 
Similarly, a system designed to merely obey the rules can be perceived as dull (who would get stuck behind a double parked vehicle?). 

This work aims at combining these two aspects. 
The interactive nature is captured by a game theoretical formulation where the players express a preference on the outcomes.
The multi-objective nature of decisions is captured by a \emph{partially ordered (posetal) preference}.
The preference on the outcomes is expressed via a hierarchy of metrics which each player can specify, allowing one to choose a clear prioritization between objectives that cannot be bargained (e.g., collision and comfort).
At the same time, players can also also express indifference between equally good alternatives (e.g., comfort and trip duration).
Borrowing the idea from the work on~\emph{minimum-violation planning}, each metric can be interpreted as a soft constraint which gets systematically violated only if inevitable.

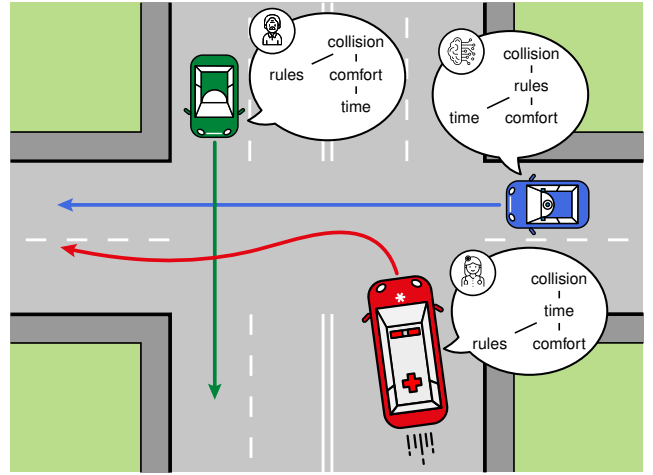
\begin{figure}[t]
    \begin{center}
    \begin{tikzpicture}
    \node at (0,0) {\includegraphics[width=0.95\linewidth]{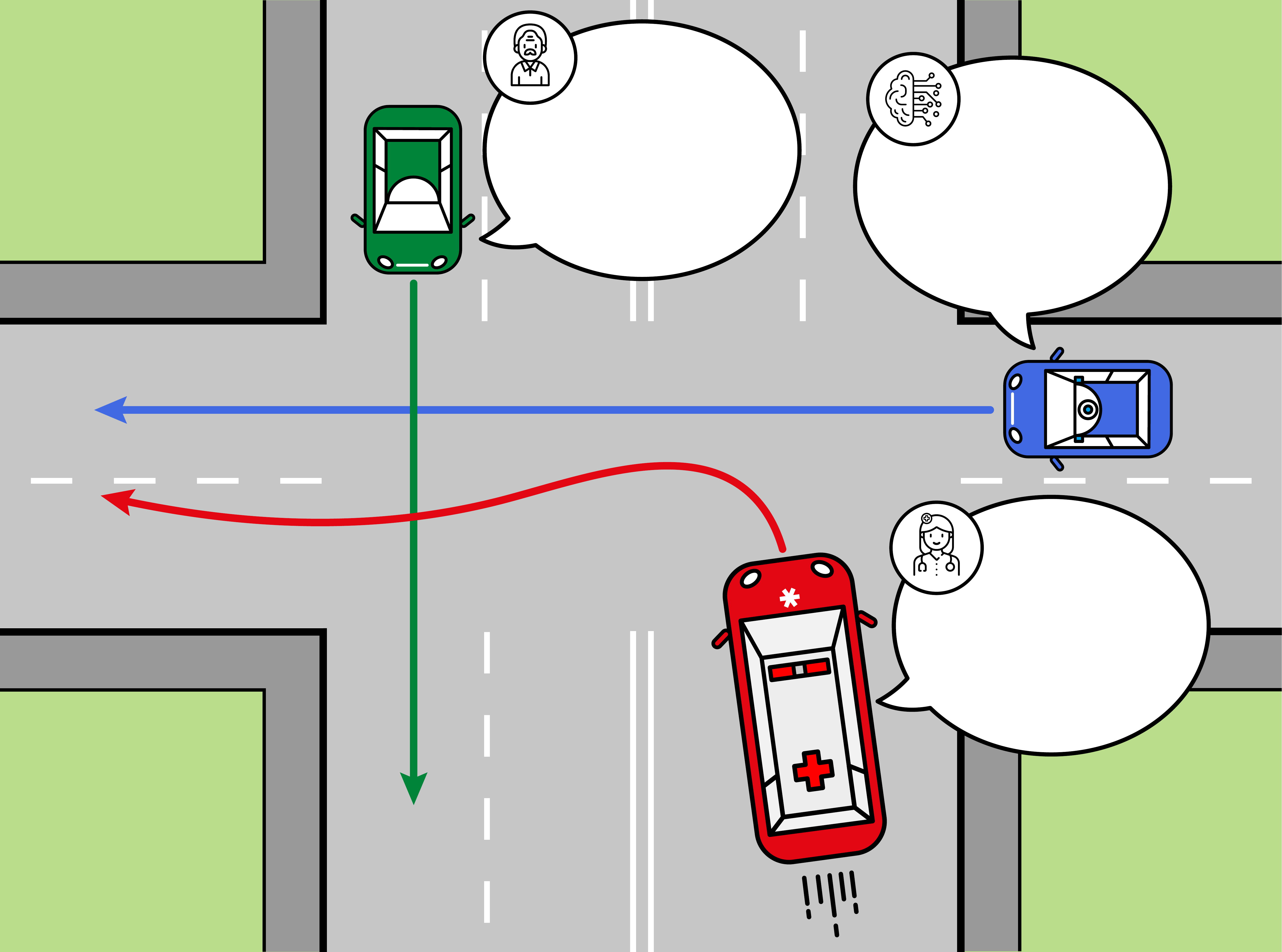}};
    \begin{scope}[scale=0.75,transform shape]
    \node at (3.15,2.7) {
    \begin{tikzcd}[row sep=small, column sep=tiny]
    &\colltext\\[-3pt]
    &\textsf{rules}\arrow[u,thick,dash]\\[-3pt]
    \epitext\arrow[ur,thick,dash]&\textsf{comfort}\arrow[u,thick,dash]
    \end{tikzcd}};
    \node at (0,2.9) {
    \begin{tikzcd}[row sep=small, column sep=tiny]
    &\colltext\\[-3pt]
    \textsf{rules}\arrow[ur,thick,dash]&\textsf{comfort}\arrow[u,thick,dash]\\[-3pt]
    &\epitext\arrow[u,thick,dash]
    \end{tikzcd}};
    \node at (3.6,-1.3) {
    \begin{tikzcd}[row sep=small, column sep=tiny]
    &\colltext\\[-3pt]
    &\epitext\arrow[u,thick,dash]\\[-3pt]
    \textsf{rules}\arrow[ur,thick,dash]&\textsf{comfort}\arrow[u,thick,dash]
    \end{tikzcd}};
    \end{scope}
    \end{tikzpicture}
    \end{center}
    
    \caption{
    In this work we look at games where each player expresses a preference as a partial order over a set of metrics. 
    This allows to model each participant with its own set of priorities.
    For instance, in the context of urban driving, everyone has collision avoidance as top priority.
    An ambulance subsequently prioritizes the minimization of travel time. Strictly following the rules of the road and guaranteeing a comfortable ride only come as tertiary objectives. 
    Conversely, an old driver, while willing to avoid collisions, is keen to respect the traffic rules and have a comfortable ride, time comes after.
    Extending standard game theoretical concepts to this new setting, allows to find equilibria solutions that are interpretable in terms of minimum violation planning. We provide sufficient conditions for equilibria to exist, and showcases the results in a detailed case study on trajectory driving games for \glspl{abk:av}.
    }
    \label{fig:mainfigure}
\end{figure}

\subsection{Related Work}
Game theoretic models aiming at describing the interactive nature of multi-robot scenarios have seen quite a revival in the last years, both looking at planning aspects of the problem~\cite{Sadigh2018,Dreves2018,Fisac2019,Fridovich-Keil2020,Fridovich-Keil2020a,Tian2020Game-TheoreticValidation} and at estimation and learning of others' cost functions~\cite{Schwarting2019a,LeCleach2021,Brito2021}. 
Notably, few examples have also been deployed on commercial \glspl{abk:av}~\cite{Ding2021EPSILON:Environments}.
However, all the proposed studies have as common denominator the usage of a scalar cost function.

Additionally, motivated by the complexity of traffic rules and the need for a transparent and interpretable system, another body of literature looked at techniques to specify objectives in a prioritized fashion~\cite{Censi2019b,Wongpiromsarn2020,Collin2020,zanardi2021udgs}, leading to practical results~\cite{BassamHelou}.

Combining these two aspects results in games where players' preferences are expressed as a binary relation on the outcomes. 
This kind of games is almost as old as game theory itself. 
Motivated by the attempt to model human decision making more accurately, already in~\cite{luce1956,Shapley1974GameUtility,Rozen2018,LeRoux2008} one can find formulations where players' preferences are non trivial binary relations (e.g., interval orders, lexicographic orders, and semiorders).
However, since these works originate from different epochs and fields, they often fail to relate to modern applications in robotics.
\subsection{Statement of Contribution}
We formulate games where each player expresses a preference as a poset of metrics, inducing a preorder on the decision space. 
As shown in~\cite{Censi2019b}, this is a practical and scalable approach for behavior specification of a robot which needs to satisfy multiple (often contrasting) objectives at the same time. 
We analyze the resulting game providing three major theoretical contributions:
\begin{itemize}
    \item We enrich the connotation of admissible equilibria by showing how they guarantee efficiency in the partial order of preferences.
    \item We provide two sufficient conditions for the existence of a pure~\gls{abk:ne} in posetal games with finite action sets.
    Interestingly, this depends on properties of the single metrics, but also on the combined preference structure of the players. 
    In addition, we motivate the conditions by introducing two examples in which a pure~\gls{abk:ne} does not exist. 
    Each example violates only one condition at a time.
    \item We show that the set of equilibria of such a game is intimately related to the operations one performs on the preference structures.
    Particularly, any refining operation of a player's preference refines the set of equilibria.   
\end{itemize}
\noindent Finally, we showcase the discovered properties in a finite trajectory driving game, instantiated on CommonRoad scenarios~\cite{Althoff2017b}.

\paragraph*{Manuscript organization}
\cref{sec:preliminaries}~provides the necessary preliminaries,  formally defining the concept of metrics, preferences, and refinement. 
\cref{sec:pos_games}~introduces \emph{Posetal Games} providing results on fairness of admissible~\gls{abk:ne}, existence of pure \gls{abk:ne}, and equilibria refinement.
These concepts are showcased in trajectory driving games in~\cref{sec:first_case_study}. 
\vspace{-0.1cm}
\section{Preliminaries}\label{sec:preliminaries}
The ensuing formalization builds on~\cite{Censi2019b,Wongpiromsarn2020}, and in general on the related work on \emph{minimum-violation} planning, where the objectives and constraints of an agent are prioritized according to a hierarchy. 
First, we define the concept of metric and model a player's preference as a prioritized order of metrics. 
Second, we consider an order among preferences which will naturally induce a refinement in the space of solutions. 
Finally, we recall desirable operations that can be applied to a preference structure, with a focus on those resulting in a \emph{refinement}. 
We assume the reader is familiar with basic facts of order theory~\cite{PriestleyIntroductionOrder}. 
\subsection{Preferences over metrics}
\begin{definition}[Metric]
\label{def:metric}
Consider a compact set~$\trajSpace$, representing a decision space.
The~$k$-th \emph{metric} is a map~$\metric^k\colon \trajSpace\to\outcomeset^k$, where~$\outcomeset^k$ is the corresponding outcome set. We assume~$\metric^k$ to be lower semi-continuous. 
\end{definition}
We will omit the superscript to indicate the union of all the metrics, that is,~$\metric = [\metric^0,\ldots,\metric^k,\ldots]$.
For the sake of simplicity, throughout this manuscript we will consider~$\outcomeset^k=\posreals$. 
Nevertheless, with the due precautions, the presented results hold for any preordered outcome set.

\begin{figure}[t]
     \begin{center}
     \begin{tikzpicture}
      \node at (0,0) {\includegraphics[width=0.9\linewidth]{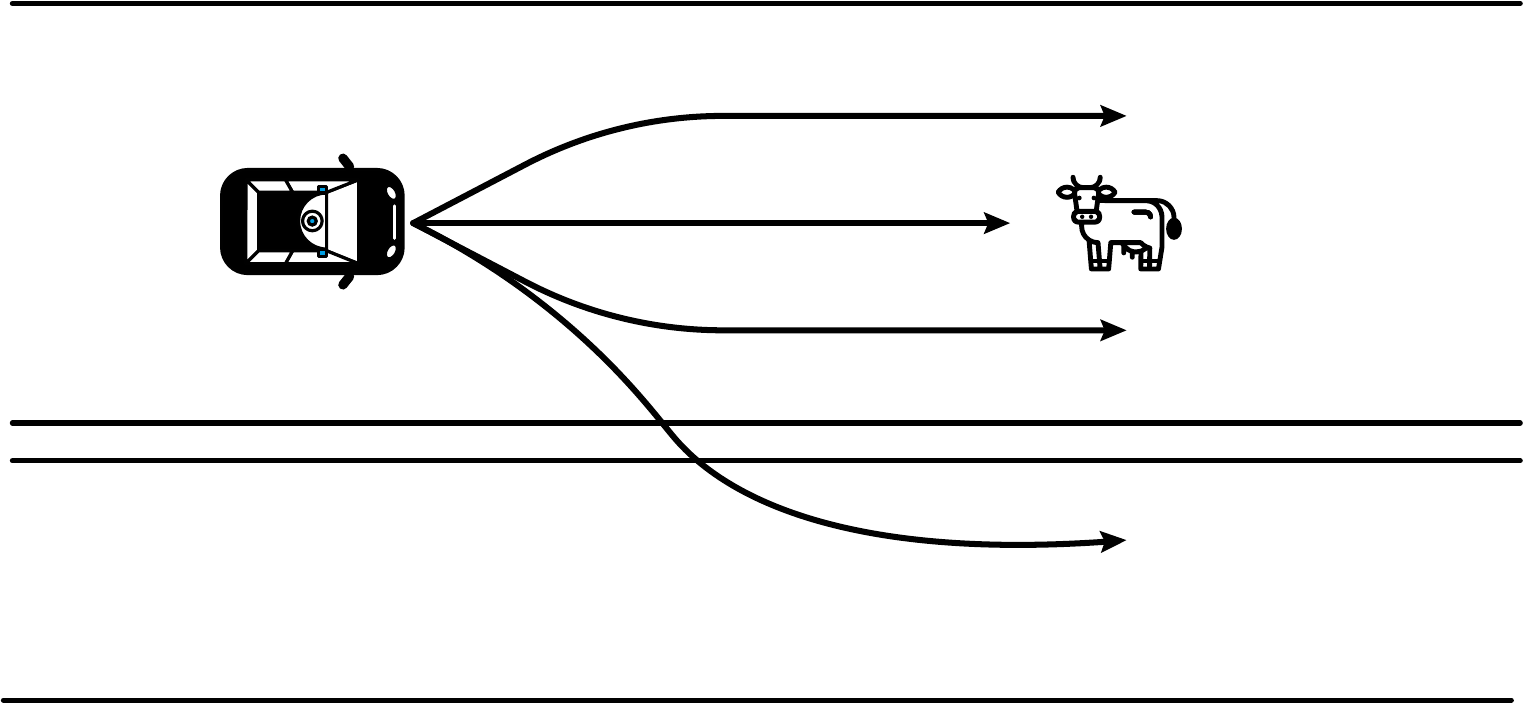}};
      \node at (0.5,1.6) {$b$};
      \node at (0.5,0.9) {$a$};
      \node at (0.5,0.25) {$c$};
      \node at (0.5,-0.75) {$d$};
\node[blockfill] at (-2.3,-1.2) {
    \begin{tikzcd}[row sep=small, column sep=tiny, every arrow/.append style={dash, thick,solid}]
\colltext&\\[-3pt]
\drivtext\arrow[u,dash,thick]&\mincltext\arrow[ul,dash,thick]
\end{tikzcd}};      
     \end{tikzpicture}
     \end{center}
     \caption{
     An \gls{abk:av} prioritizing the minimization of its collision cost (e.g. kinetic energy transfer) over a severe violation of traffic rules (e.g. cumulative time beyond the double lines) and a minimum clearance cost (e.g. safety distance violation). 
In this setting, trajectories~$b$,~$c$, and~$d$ are preferred over trajectory~$a$.
Even though~$b$ and~$c$ are two different trajectories, they evaluate to be \emph{indifferent} since they lead to the \emph{same} outcomes. Both share the same collision, area violation, and clearance values.
Finally, trajectory~$d$ has a worse area violation, but a better clearance than~$b,c$. It is therefore \emph{uncomparable} with respect to~$b,c$, and constitutes a so-called Pareto front with them.}
     \label{fig:comparison}
\end{figure}
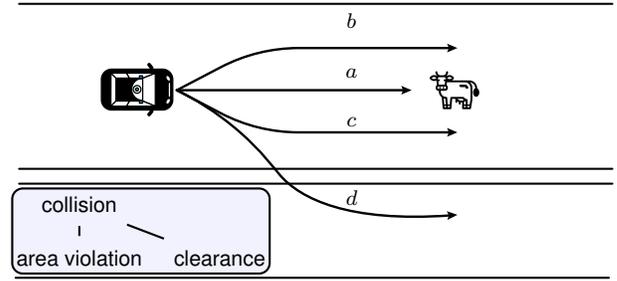

\begin{definition}[Preference]\label{def:preference}
Given a set of metrics~$\metricsSet=\{\metric^1,\ldots,\metric^n\}$. A \emph{preference}~$\preference$ is specified as a partial order over~$\metricsSet$, i.e.,~$\preference = \tup{\metricsSet,\posleq}$. 
\end{definition}
A simple preference from the driving domain is shown in~\cref{fig:comparison}.\footnote{We represent preferences via Hasse diagrams. 
In a Hasse diagram for poset~$\preference$, one writes~$\metric^1$ below~$\metric^2$ and connect them with a line if~$\metric^1\posleq_\preference \metric^2$. Relations arising from reflexivity and transitivity are omitted.}
As observed in~\cite{Censi2019b}, such a preference (\cref{def:preference}) naturally induces a pre-order on the outcome set~$\outcomeset = \Pi_k \outcomeset^k$, denoted by~$\tup{\outcomeset,\precsim}$. More importantly, a preorder is induced also over the decision space.
\begin{lemma}[Preorder induced on decision space]
\label{lem:preorder_decisionspace}
A preference~$\preference$ (\cref{def:preference}) induces a preorder on the decision space~$\tup{\trajSpace,\precsim_\trajSpace}$, where
\begin{equation}
    \traj \precsim_\trajSpace \traj^\prime \Leftrightarrow \metric(\traj)\precsim_\outcomeset \metric(\traj^\prime).
\end{equation}
\end{lemma}
\ifthenelse{\boolean{proofs}}{\begin{proof}[Proof\ifthenelse{\boolean{proofs}}{}{
of \cref{lem:preorder_decisionspace}}]
Reflexivity is clear. 
Suppose~$\traj\precsim_\trajset \traj'$ and~$\traj'\precsim_\trajset \traj''$, i.e.,~$\metric(\traj)\precsim_\outcomeset \metric(\traj')$ and~$\metric(\traj')\precsim_\outcomeset \metric(\traj'')$. Since~$\tup{\outcomeset,\precsim_\outcomeset}$ is a preorder,~$\metric(\traj)\precsim_\outcomeset \metric(\traj'')$, and hence~$\traj\precsim_\trajset \traj''$.
\end{proof}}{}
Importantly, when comparing two elements in a preordered set, one can have four possible outcomes: the \emph{first} is \emph{preferred}, the \emph{second} is \emph{preferred}, the two are \emph{uncomparable}, or, they are \emph{indifferent}.
This concept is exemplified when comparing trajectories of an \gls{abk:av} in~\cref{fig:comparison}.

\subsection{Order on preferences and refining operations}
Given the set of all possible metrics~$\allmetrics$, we denote by~$\pre{\allmetrics}$ the set of \emph{all} preorders over~$\allmetrics$. Note that a preference could be in general defined only on a subset~$\metricsSet\subseteq\allmetrics$ of all the metrics. 
Technically, this collapses the outcome set for the metrics which do not appear in the preference to an equivalence class: the player is \emph{indifferent} to any value such metrics might assume. 

Preferences themselves can be ordered via a preorder.
\begin{definition}[Preorder of Preferences]
\label{def:preorder_preferences}
We define a preorder of preferences~$\tup{\pre{\allmetrics},\precsim}$ as follows.
Given any two preferences~$\preference =\tup{\metricsSet,\posleq},\, \preference^\prime=\tup{\metricsSet',\posleq^\prime}\in \pre{\allmetrics}$, one has:
\begin{equation}\label{eq:pre_pre}
\preference \precsim \preference^\prime \Leftrightarrow 
 \tup{\outcomeset,\prec}\subseteq \tup{\outcomeset,\prec^\prime},
\end{equation}
meaning that~$\preference$ relates to~$\preference^\prime$ iff, when a pair of outcomes~$\tup{\outcome_1,\outcome_2}$ is in strict relation (either \emph{first-} or \emph{second-preferred}) according to~$\preference$, the same relation must hold for~$\preference^\prime$.
\end{definition}
\begin{lemma}
\label{lem:preorder_pref}
\cref{def:preorder_preferences} indeed defines a preorder.
\end{lemma}
\ifthenelse{\boolean{proofs}}{

\begin{proof}[Proof\ifthenelse{\boolean{proofs}}{}{
of \cref{lem:preorder_pref}}]
Clearly, one has~$\tup{\outcomeset,\prec}\subseteq \tup{\outcomeset,\prec}$ (reflexivity). 
Furthermore, if~$\tup{\outcomeset,\prec}\subseteq \tup{\outcomeset,\prec'}$ and~$\tup{\outcomeset,\prec'}\subseteq \tup{\outcomeset,\prec''}$, then~$\tup{\outcomeset,\prec}\subseteq \tup{\outcomeset,\prec''}$ (transitivity). 
\end{proof}}{}
We can now leverage \cref{def:preorder_preferences} to define the concept of \emph{preference refinement}.
\begin{definition}[Preference refinement]\label{def:refinement}
Let~$\preference, \preference' \in \pre{\allmetrics}$. 
Preference~$\preference'$ is a \emph{refinement} of~$\preference$ if $\preference \precsim \preference'$ as by~\cref{def:preorder_preferences}.
\end{definition}
\noindent This notion of refinement is analogous to~\cite[Def. 18]{Censi2019b}.

\subsection{Refining operations on preferences}
We now look at preference-refining operations.
Without loss of generality, we define such operations directly on preferences.
Due to the induced preorders on outcomes and decision space (\cref{lem:preorder_decisionspace}), the refinement propagates. 

\paragraph{Priority refinement}
This operation corresponds to ``adding an edge'' to the graph representing the preference. 
\begin{definition}[Priority refinement operation]
\label{def:priority_ref_op}
Consider a preference~$\preference=\tup{\metricsSet,\posleq}$. 
A \emph{priority refinement operation} on~$\preference$ is any operation~$z\colon \pre{\metricsSet}\to \pre{\metricsSet}$ such that~$z(\preference)$ refines~$\preference$.
\end{definition}

\paragraph{Aggregation}
This operation allows one to condense two metrics of the preference into a single one.
\begin{definition}[Aggregation]
\label{def:aggregation_operation}
Given a preference~$\preference=\tup{\metricsSet,\posleq}$ and two uncomparable metrics~$\metric^1,\metric^2\in \metricsSet$, 
an \emph{aggregation} for~$\metric^1,\metric^2$ is given by~$\aggregation_\alpha\colon\metricsSet\times \metricsSet\to \metricsSet\times \metricsSet$,~$\aggregation_\alpha(\preference)\coloneqq\preference'$, with
\begin{equation*}
    \preference'=S_1\cup S_2 \cup \{\tup{\metric,\metric'}\in \preference \mid \metric,\metric' \in \metricsSet \backslash \{\metric^1,\metric^2\}\},
\end{equation*}
\begin{equation*}
    \begin{aligned}
    S_i&=\bigcup_{\tup{\metric^i,\metric}\in \preference}\tup{\alpha(\metric^1,\metric^2),\metric}\cup 
    \bigcup_{\tup{\metric,\metric^i}\in \preference}\tup{\metric,\alpha(\metric^1,\metric^2)},
    \end{aligned}
\end{equation*}
where~$\alpha$ is an embedding of the product poset over~$\metric^1\times \metric^2$ into~$\posreals$.
\end{definition}

\begin{remark}
$\alpha$ needs to be a strictly monotone map in both arguments.
Allowed choices include linear combinations with positive coefficients:~$\alpha(\metric^1,\metric^2)=a\metric^1+b\metric^2$,~$a,b\in \mathbb{R}_{>0}$.
This construction is similar in purpose to~\cite[Def. 16]{Censi2019b}, but we aggregate uncomparable (and not indifferent) metrics.
\end{remark}

\begin{figure}[t]
    \centering
    \begin{tikzpicture}
\begin{scope}[scale=0.9, transform shape]
\node[blockfill, inner sep=-3pt] at (0,0) (first) {
\begin{tikzcd}[row sep=small, column sep=tiny, every arrow/.append style={dash, thick, solid}]
&\metric^1&\\[-3pt]
\metric^2 \arrow[ur,dash,thick]&&\metric^3\arrow[ul,dash,thick]
\end{tikzcd}
};
\node[above left=0.5cm and 0.5cm of first,blockfill, inner sep=-3pt] (third) {
\begin{tikzcd}[row sep=small, column sep=tiny, every arrow/.append style={dash, thick,solid}]
\metric^1\\[-3pt]
\metric^3\arrow[u,dash,thick]\\[-3pt]
\metric^2\arrow[u,dash,thick]
\end{tikzcd}
};
\node[above right=0.5cm and 0.5cm of first,blockfill, inner sep=-2pt] (second) {
\begin{tikzcd}[row sep=small, column sep=tiny, every arrow/.append style={dash, thick, solid}]
\metric^1\\
\alpha(\metric^2,\metric^3)\arrow[u,dash,thick]
\end{tikzcd}
};

\node[above =0.5cm of first,blockfill, inner sep=-2pt] (fourth) {
\begin{tikzcd}[row sep=tiny, column sep=tiny, every arrow/.append style={dash, thick, solid}]
&\metric^1&\\
\metric^2 \arrow[ur,dash,thick]&&\metric^3\arrow[ul,dash,thick]\\
&\metric^4\arrow[ur,dash,thick]\arrow[ul,dash,thick]&
\end{tikzcd}
};
\draw[-Triangle, thick] (first) to[bend left=20] node[pos=0.5,left]{\begin{tabular}{c}priority \\refinement\end{tabular}}(third.south) ;
\draw[-Triangle, thick] (first) to[bend right=20] node[pos=0.5,right]{ aggregation} (second.south);
\draw[-Triangle, thick] (first) -- node[pos=0.5,right]{augmentation} (fourth.south);
\end{scope}
\end{tikzpicture}
    \caption{We represent preferences using Hasse diagrams. Priority refinement (\cref{def:priority_ref_op}, adding an edge between~$\metric^2$ and~$\metric^3$), aggregation (\cref{def:aggregation_operation}, aggregating~$\metric^2$ and~$\metric^3$ via~$\alpha$), and augmentation (\cref{def:augm_op}, augmenting with~$\metric^4$).}
    \label{fig:refinement_ops}
\end{figure}
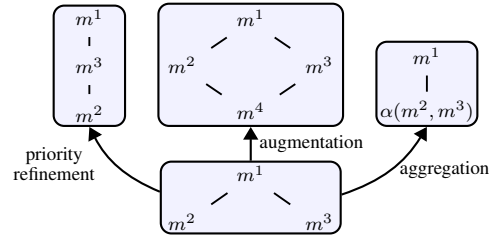

\paragraph{Augmentation}
This operation allows one to add a new metric at the lowest level of priority. 
Note that as discussed in~\cite{Censi2019b}, adding a rule can be a more general operation. 
Yet, adding a rule at the ``bottom'' results in a refinement; completely different preferences could otherwise arise. 
Therefore, in the following, we refer to \emph{augmentation} in this sense only.
\begin{definition}[Metric augmentation operation]
\label{def:augm_op}
Consider a preference~$\preference=\tup{\metricsSet,\posleq}$. 
\emph{Metric augmentation} consists of augmenting~$\metricsSet$ to~$\metricsSet'=\metricsSet\cup \{\metric'\}\subseteq\allmetrics$ and defining a preference~$\preference'=\tup{\metricsSet',\posleq'}$ such that~$\metric'\posleq' \metric$ for all~$\metric\in \metricsSet$.
\end{definition}
\noindent \cref{fig:refinement_ops} exemplifies the three operations (\cref{def:priority_ref_op,def:aggregation_operation,def:augm_op}).

\begin{lemma}
\label{lem:operations_refine}
Consider a preference~$\preference=\tup{\metricsSet,\posleq}$. Applying any of the operations (priority refinement, aggregation, and augmentation) to~$\preference$ results in a preference~$\preference'$ which refines~$\preference$.
\end{lemma}
Technically, \cref{lem:operations_refine} states that refinement operations are monotone maps~$\pre{\allmetrics}\to \pre{\allmetrics}$.

\ifthenelse{\boolean{proofs}}{}{
\input{sections/proofs/operations_refine}}

\section{Games With Posetal Preferences}
\label{sec:pos_games}
In the last decades, games where players express a general preference on outcomes have been investigated~\cite{Rozen2018,LeRoux2008,luce1956}. 
Yet, these studies do not instantiate these concepts in engineering frameworks where, for instance, expressing the preference directly on outcomes or on decision spaces is impractical. 

We now formally consider games where players express a preference over the outcomes via a prioritized hierarchy of objectives and constraints (the posetal preference). 
\begin{definition}[Game with Posetal Preferences]
\label{def:posgame}
A \emph{\gls{abk:gpop}} (in short, posetal game) is specified as follows.
\begin{itemize}
    \item There is a \emph{finite} set of players~$\players$;
    \item Each player possesses a compact decision space, denoted by~$\trajset_i$.
    The joint decision space is~$\trajset=\prod_i \trajset_i$,
    thus~$\traj=\tup{\traj_i,\traj_{-i}}$ denotes a joint action profile.
    \item Given the action profile for the players, an outcome of the game for each player~$i\in \players$ is obtained via a deterministic\footnote{While extensions to stochastic frameworks have been proposed in~\cite{Rozen2018}, in this work we consider pure actions and deterministic metrics.} metric function, mapping joint decisions to the outcome space~$\metric_i\colon \trajset \to \outcomeset_i$ (\cref{def:metric}).
    \item Each player~$i\in \players$ specifies a preference~$\preference_i$ (\cref{def:preference}). It follows (\cref{lem:preorder_decisionspace}) that the induced preorder on decisions~$\tup{\trajset_i,\precsim_{\trajset_i}}$ is:
    \begin{equation*}
        \traj_i\precsim_{\trajset_i} \traj_i' \Leftrightarrow \metric_i(\argtuple{\traj_i}{\traj_{-i}}) \precsim_{\outcomeset_i}\metric_i(\argtuple{\traj_i'}{\traj_{-i}}) \ \forall \traj_{-i}\in \trajset_{-i}.
    \end{equation*}
\end{itemize}
\end{definition}

Given the setup of a~\gls{abk:gpop}, standard game theory concepts follow naturally.  
To characterize \emph{rational} players and the solutions of a game, we first consider \gls{abk:br}.

\begin{definition}[\gls{abk:br} for \glspl{abk:gpop}]
\label{def:br}
The \gls{abk:br} for Player~$i\in \players$ is a set-valued map $\br_i:\trajset_{-i}\to \trajset_i$. A \emph{strict} and \emph{weak} version, respectively ~$\brstrict_i,\brweak_i$,  are given by
\begin{equation*}
    \begin{aligned}
    \brstrict_i\colon %
    \traj_{-i}&\mapsto \{ \traj_i\in \trajset_i \mid \metric_i(\traj)\prec_{\outcomeset_i}\metric_i(\argtuple{\traj_i'}{\traj_{-i}}) \ \forall \traj_i'\in \trajset_i\},\\
    \brweak_i\colon %
    \traj_{-i}&\mapsto \{ \traj_i\in \trajset_i \mid \metric_i(\argtuple{\traj_i'}{\traj_{-i}})\nprec_{\outcomeset_i}\metric_i(\traj) \ \forall \traj_i'\in \trajset_i\}.
    \end{aligned}
\end{equation*}
\end{definition}
Note that given a non-empty set of alternatives,~$\brweak_i(\traj_{-i})$ is always non-empty. Indeed, for every player~$i\in \players$, the weak \gls{abk:br} map corresponds to finding the \emph{minimal elements} of the preorder induced on the decision space (\cref{lem:preorder_decisionspace}).

The \gls{abk:br} map definition allows one to formally define the notion of \gls{abk:ne} of \glspl{abk:gpop}.

\begin{definition}[\gls{abk:ne} of \glspl{abk:gpop}]
\label{def:nasheq}
Let~$\game$ be a \gls{abk:gpop}. 
A strategy profile~$\traj \in \trajset=\prod_{i\in \players}\trajset_i$ is a \gls{abk:ne} iff it is a \gls{abk:br} for all players. From the best-responses we distinguish \emph{strict} and \emph{weak} \gls{abk:ne}:
\begin{equation*}
\begin{aligned}
   \traj_{i} &\in \br^{\prec\;[\preleq]}_i (\traj_{-i})\quad\forall i\in\players \quad \text{strict [weak]},\\
\end{aligned}
\end{equation*}
and denote by~$\nasheq^{\prec\;[\preleq]}(\game)$ the set of strict [weak] \gls{abk:ne} of~$\game$.
\end{definition}

\begin{remark}[Semantics of \gls{abk:ne}]
A \emph{strong} \gls{abk:ne} consists of a strategy profile where the choice of every player is \emph{first-preferred} to \emph{all} the alternative ones. The notion of \emph{weak} \gls{abk:ne} captures the cases in which for some players a strategy can be \emph{uncomparable} or \emph{indifferent} to some of the alternatives, yet there is none which is strictly better. Clearly,~$\neqstrong(\game)\subseteq \neqweak(\game)$.
\end{remark}

Finally, as in standard game-theory, one can consider \emph{dominating} \gls{abk:ne}, called \emph{admissible}.
\begin{definition}[Admissible \gls{abk:ne}]
\label{def:admissible_ne}
Consider a game~$\game$ with a set of \gls{abk:ne}~$\neqweak(\game)$. The set of \emph{admissible \gls{abk:ne}} is given by
\begin{equation*}
   \neqweakadm(\game)\coloneqq \{\traj \in \neqweak(\game)\mid \metric(\traj')\nprec_{\outcomeset}\metric(\traj) \ \forall \traj'\in \neqweak(\game)\}.
\end{equation*}
\end{definition}
\begin{remark}
Clearly,~$\neqweakadm(\game)\subseteq \neqweak(\game)$.
\end{remark}
\subsection{Rank of Solutions}
In the setting of \glspl{abk:gpop}, admissible \gls{abk:ne} not only are a proxy for \gls{abk:ne} efficiency, but also guarantee efficiency on the hierarchy of priorities (\cref{lem:admissible_rank}).
To formalize this concept, we first define the notion of \emph{rank}.
\begin{definition}[Rank of a metric]\label{def:rankofmetric}
Given a preference~$\preference$, each metric~$\metric$ has \emph{rank}~
   $\rankk(\metric)\coloneqq \height{Q}$,~$Q=\ \upit \{\metric\}$,
where~$\height{Q}$ represents the maximum cardinality of a chain in~$Q$ and~$\upit$ represents the upper closure operator.
\end{definition}
Loosely speaking, we can see the Hasse diagram of a preference as a  downward-directed graph. Then, the rank of a metric corresponds to the length of the longest path (chain) from any root (top) to the metric.

We can now extend the notion of rank to players' actions.
Here, the rank of an action describes the \emph{critical priority} that a player can reach without violating any metric of higher order.\footnote{We see this as an analogy to active constraints in constrained optimization.}

\begin{definition}[Rank of an action]
\label{def:rank_decision}
Consider a \gls{abk:gpop}~$\game$. 
The \emph{rank of action}~$\traj\in\trajset$ for player~$i\in \players$ with preference~$\preference_i=\tup{\metricsSet,\posleq}$ is denoted~$\rankdec_i(\traj)$, and is given by

\begin{equation*}
    \begin{cases}
    \rankk(\metric_i^k(\traj)), & \metric_i^k(\traj)\neq \Min\tup{\outcomeset^k,\precsim} \wedge\\
    &(\forall \metric_i^l(\traj)\posgeq \metric_i^k(\traj): \metric_i^l(\traj)=\Min\tup{\outcomeset^l,\precsim}),\\
    \rankk(\Min \preference_i(\traj)), & \text{otherwise}\\
    \end{cases}
\end{equation*}
\end{definition}
\cref{def:rank_decision} can be interpreted as follows. The rank of a joint action from the viewpoint of a specific player is a proxy for the satisfaction level of the outcomes achieved through the action (measured as the rank of highest non-minimized metrics in the player's preference poset). 
If all metrics in the preference structure are minimized, the rank is simply the rank of the minima in the preference poset.

\cref{def:rank_decision} can be extended to account for multiple players.
\begin{definition}[Common rank of a strategy profile]
\label{def:rank_ne}
Consider a \gls{abk:gpop}~$\game$ with~$n$ players.
The \emph{common rank} of a strategy profile~$\traj\in \trajset$ is:
    $\rankne(\traj)\coloneqq \min\{\rankdec_1(\traj),\ldots, \rankdec_n(\traj)\}.$
\end{definition}

\begin{lemma}
\label{lem:admissible_rank}
Consider a \gls{abk:gpop}~$\game$ with a set of players~$\agents$, admitting a set of \gls{abk:ne}~$\neqweak(\game)$.
Consider~$\trajne \in \neqweak(\game)\backslash \neqweakadm(\game)$ and~$\trajne{'}\in \neqweakadm(\game)$.
It holds
\begin{equation}
\label{eq:admissible_rank}
\rankdec_i(\trajne)\leq \rankdec_i(\trajne{'}) \quad \forall i\in \agents.
\end{equation}
Note that \cref{eq:admissible_rank} implies~$\rankne(\trajne)\leq \rankne(\trajne{'})$.
\end{lemma}

\ifthenelse{\boolean{proofs}}{
\begin{proof}
The result follows directly from \cref{def:br}, \cref{def:nasheq}, \cref{def:admissible_ne}, and \cref{def:rank_decision}.
\end{proof}}{}

\begin{remark}
In other words, admissible \gls{abk:ne} not only guarantee dominating payoffs among \gls{abk:ne} as per \cref{def:admissible_ne}, but also guarantee a same or higher rank (i.e., that the critical metric has same or lower priority) with respect to any non-admissible \gls{abk:ne}.
\end{remark}

\subsection{On the existence of~\gls{abk:ne} in \gls{abk:gpop}s}

The seminal work of Nash~\cite{Nash1950} that proved the existence of equilibria for finite games in mixed strategies has been extended to games with ordered preferences. 
A review is provided by~\cite{Rozen2018}.
Nevertheless, games where the existence of \gls{abk:ne} can be guaranteed in \emph{pure} actions are often desirable for computational reasons in engineering applications.

In the following, we provide sufficient conditions which guarantee the existence of pure~\gls{abk:ne} for~\glspl{abk:gpop} with a discrete action space.
We translate the idea of the ``players \emph{not} having contrasting coupled objectives'' to conditions on single metrics and combined preferences of the players. 
These allow to define a surrogate potential for the game. Then, the existence of pure \gls{abk:ne} follows via standard game theory results~\cite{Monderer1996}.

\paragraph{Jointly Communal Objectives}
We consider the idea of players not having contrasting objectives by extending the idea of communal metrics~\cite[Def. 4]{zanardi2021udgs} to arbitrary pairs of metrics.
The idea is that if a player improves on a metric through a unilateral deviation in strategy, and the others do not collectively deteriorate their performance in a given metric, the two metrics are called \emph{jointly communal}.

\begin{definition}[Jointly communal metrics]
\label{def:jointly_communal}
Consider a \gls{abk:gpop} with players~$\agents$.
We say that metric~$\metric^k,\metric^l$ are \emph{jointly communal} if and only if, for all~$i\in \agents$,~$\traj,\traj'\in \trajset$ it holds
\begin{equation}
\label{eq:joint_communal}
    \metric_i^k(\traj')-\metric_i^k(\traj)<0 \Rightarrow %
    \sum_{j\in \agents^l}
    (\metric_j^l(\traj')-\metric_j^l(\traj))\leq0,
\end{equation}
where~$\agents^l\subseteq \agents$ represents the subset of agents featuring the~$l$-th metric in their preference structure.
\end{definition}
Note that~\cref{eq:joint_communal} is not restrictive for many common cost functions used in mobile robotics, including, among others, \emph{collision} costs, \emph{clearance} objectives (i.e., keeping a \emph{minimum safety distance} from obstacles)~\cite{zanardi2021udgs}.
Moreover, any pair of personal metrics (i.e., metrics whose outcome depends on the single player strategy only) is trivially jointly communal.

To guarantee existence of \gls{abk:ne}, \cref{def:jointly_communal} needs to hold for all uncomparable metrics in a game.
Formally, we require that:

\begin{condition}
\label{cond:join_all}
Let~$\joinofprefs\coloneqq \bigcup_{i\in \agents} \preference_i$. Then, all incomparable metrics in~$\joinofprefs$ are pairwise jointly communal.
\end{condition}
The relation~$\joinofprefs$ is the union of the players' preferences.
A visual example is provided in~\cref{fig:jointprefreduction}.

\begin{figure}[t]
    \centering
    \begin{tikzpicture}
\begin{scope}[scale=0.9, transform shape]
\node[blockfill](first) at (-2,0) {\begin{tikzcd}[row sep=small, column sep=tiny]
\jointmetric_1^1&\\[-3pt]
\metric_1^2\arrow[u,dash,solid,thick]&\metric_1^3\arrow[ul,dash,solid,thick]\\[-3pt]
\jointmetric_1^4\arrow[u,dash,solid,thick]&\jointmetric_1^5\arrow[u,dash,solid,thick]
\end{tikzcd}};
\node[blockfill, right=0.5cm of first] (second) {
\begin{tikzcd}[row sep=small, column sep=tiny]
\jointmetric_2^1\\[-3pt]
\jointmetric_2^4 \arrow[u,dash,solid,thick]\\[-3pt]
\metric_2^6\arrow[u,dash,solid,thick]
\end{tikzcd}
};
\node[blockfill, right=0.5cm of second] (fifth) {
\begin{tikzcd}[row sep=small, column sep=tiny]
\jointmetric_1^1&&\jointmetric_2^1\\[-3pt]
\metric_1^2\arrow[u,dash,solid,thick]&\metric_1^3\arrow[ul,dash,solid,thick]&\jointmetric_2^4 \arrow[u,dash,solid,thick]\\[-3pt]
\jointmetric_1^4\arrow[u,dash,solid,thick]&\jointmetric_1^5\arrow[u,dash,solid,thick]&\metric_2^6\arrow[u,dash,solid,thick]
\end{tikzcd}
};

\node[above=-0.1cm of first.north] {$\preference_1$};
\node[above=-0.1cm of second.north] {$\preference_2$};
\node[above=-0.1cm of fifth.north] {$\joinofprefs=\preference_1\cup \preference_2$};
\end{scope}
\end{tikzpicture}
   \caption{Given relations~$\preference_1,\preference_2$ one can compute their union~$\joinofprefs$. }
    \label{fig:jointprefreduction}
    \vspace{-0.5cm}
\end{figure}

Two comments are in order.
First, if all~$\metricsSet_i$ are disjoint,~$\joinofprefs$ is clearly a poset over~$\bigcup_{i\in \agents}\metricsSet_i$.
Second, in case~$\metricsSet_i=\metricsSet_j$ for all~$i,j\in \agents$,~$\joinofprefs$ corresponds to the \emph{join} (least upper bound)~$\bigvee_{i\in \agents}\preference_i$ in~$\pre{\metricsSet}$.

\paragraph{Consistent preferences}
The complexity of allowing arbitrarily prioritized metrics requires the introduction of a second condition, now on the preferences of the players.

To ease the reading, in the following we denote with~$\jointmetric\in\jointmetricSet$ the \emph{joint} metrics, opposed to the personal ones, as the metrics that depend on the actions and states of many players~\cite{zanardi2021udgs}.
\reviewed{For instance, the time taken by an agent to cross an intersection can be computed as a function of its own trajectory only. Instead, a joint metric such as minimum clearance or collision energy is by definition a function of two or more trajectories.}
\begin{condition}
\label{cond:jointly}
Given a \gls{abk:gpop} it holds, for all~$i\in \agents$:
\begin{equation}
\label{eq:second_condition}
    \jointmetric_{i}^k\posleq_{\preference_i}\jointmetric_i^l \Rightarrow \nexists j\in \agents\backslash\{i\} \colon \jointmetric_j^l \posleq_{\preference_j} \jointmetric_j^k.
\end{equation}
\end{condition}

\paragraph{Existence of pure NE for \glspl{abk:gpop}}
\begin{theorem}
\label{thm:existence_ne}
A \gls{abk:gpop} with finite action sets satisfying~\cref{cond:jointly,cond:join_all} admits a pure~\gls{abk:ne}.
\end{theorem}

\ifthenelse{\boolean{proofs}}{
\begin{proof}
We organize the proof as follows. 
First, we show how the satisfaction of \cref{cond:jointly,cond:join_all} allows one to construct a posetal potential function~$\potential$.
Second, we show that the minimum of~$\potential$ is a pure \gls{abk:ne}.
Consider a \gls{abk:gpop} with agents~$\agents$, each characterized by a preference~$\preference_i$,~$i\in \agents$.

\noindent 1) The satisfaction of \cref{cond:jointly} guarantees that~$\joinofprefs$ is a poset.\footnote{Without loss of generality we can consider players with preferences on disjoint metric sets.}
We now construct a partially ordered potential~$\potential$ as follows. 
Leveraging \cref{def:aggregation_operation}, starting from~$\joinofprefs$, we sequentially aggregate all joint metrics~$\jointmetric_i^k$ for all players~$i\in \agents^k$.
Note that sequential aggregation is well-defined given~\cref{cond:join_all}.
An illustrative example is reported in \cref{fig:potential_example}, where we aggregate via sum.

Given~\cref{cond:jointly,cond:join_all},~$\potential$ is a poset.
In the following, we denote its carrier set by~$\metricsSet$ and~$\metricsSet\backslash \jointmetricSet$ by~$\metricsSetComp$.

\noindent Now, consider a unilateral deviation in strategy~$\traj_i\to \traj_i'$ for player~$i\in \agents$ (i.e., a strategy switch~$\traj\to \traj'$, where~$\traj=\tup{\traj_i,\traj_{-i}}$ and~$\traj'=\tup{\traj_i',\traj_{-i}}$ such that~$\metric_i(\traj')\prec_{\outcomeset_i}\metric_i(\traj)$.
We need to show~$\potential(\traj')\prec \potential(\traj)$.
We have two cases.
a) If Player~$i$ improved due to an improvement on a joint metric, i.e.,~$\jointmetric_i^l(\traj')< \jointmetric_i^l(\traj)$,~$\jointmetric_i^l\in \jointmetricSet$, \cref{cond:jointly,cond:join_all} impose improvement of~$\potential$ in the component relative to the aggregation of the~$l$-th joint metric.
Furthermore, they guarantee other components relative to joint metrics of higher priority do not deteriorate~$\potential$. 
Clearly, such an improvement does not deteriorate any personal metric in~$\potential$ for the other players.
Hence, in this case~$\potential(\traj')\prec \potential(\traj)$.
b) If, instead, Player~$i$ improved due to a personal metric, i.e.,~$\metric_i^l(\traj')< \metric_i^l(\traj)$,~$\metric_i^l\in \metricsSetComp$, clearly the corresponding component in~$\potential$ will improve.
Such an improvement could deteriorate a joint metric for Player~$i$ and other players, but the form of~$\potential$ and \cref{cond:join_all,cond:jointly} guarantee that the interested joint components of~$\potential$ will be dominated by~$\metric^l$ in priority for Player~$i$ (the case in which they are higher/uncomparable in priority was covered in a)).
Furthermore, improvement on a personal metric for Player~$i$ cannot deteriorate personal metrics of other players (\cref{cond:join_all}).
Finally, note that improvements happening both due to joint and personal metrics simultaneously are captured by the composition of the above cases.
Therefore,~$\potential(\traj')\prec \potential(\traj)$,~$\potential$ is a valid potential function, and a \gls{abk:gpop} satisfying \cref{cond:jointly,cond:join_all} is a posetal potential game.

\noindent 2) As for standard game theory~\cite{Hespanha2017}, let $\potential(\trajne)$ be a minimal element of the potential, by definition of potential nobody can unilaterally change strategy receiving a strictly better payoff otherwise $\potential(\trajne)$ would not be a minimal element of the poset. Thus $\trajne$ is a weak~\gls{abk:ne}. 
The same reasoning applies for the strict counterpart assuming~$\potential(\trajne)$ is a minimum element.
Note that the existence of such potential allows one to additionally prove the convergence of widely used algorithms, such as iterated strictly-better response schemes as in~\cite{zanardi2021udgs}.
\end{proof}
}{}

\begin{figure}[t]
\begin{center}
\begin{tikzpicture}
\begin{scope}[scale=0.9, transform shape]
\node[blockfill](first) at (-2.5,0) {\begin{tikzcd}[row sep=small, column sep=tiny]
&\jointmetric_1^1\\[-3pt]
\jointmetric_1^2\arrow[ur,dash,solid,thick]&\metric_1^4\arrow[u,dash,solid,thick]\\[+15pt]
\end{tikzcd}};
\node[blockfill, right=0.15cm of first](second) {\begin{tikzcd}[row sep=small, column sep=tiny]
\jointmetric_2^1\\[-3pt]
\metric_2^3\arrow[u,dash,solid,thick]\\[-3pt]
\metric_2^5\arrow[u,dash,solid,thick]
\end{tikzcd}};
\node[blockfill, right=0.15cm of second](third){\begin{tikzcd}[row sep=small, column sep=tiny]
&\jointmetric_1^1&\jointmetric_2^1\\[-3pt]
\jointmetric_1^2\arrow[ur,dash,solid,thick]&\metric_1^4\arrow[u,dash,solid,thick]&\metric_2^3\arrow[u,dash,solid,thick]\\[-3pt]
&&\metric_2^5\arrow[u,dash,solid,thick]
\end{tikzcd}};
\node[blockfill, right=0.15cm of third](fourth){\begin{tikzcd}[row sep=small, column sep=small]
&[+10pt]\jointmetric_1^1+\jointmetric_2^1\\[-3pt]
\jointmetric_1^2\arrow[ur,dash,solid,thick]&\metric_2^3\arrow[u,dash,solid,thick]\\[-3pt]
\metric_1^4\arrow[uur,dash,solid,thick]&\metric_2^5\arrow[u,dash,solid,thick]
\end{tikzcd}};
\node[above=-0.1cm of first.north] {$\preference_1$};
\node[above=-0.1cm of second.north] {$\preference_2$};
\node[above=-0.1cm of third.north] {$\joinofprefs=\preference_1 \cup \preference_2$};
\node[above=-0.1cm of fourth.north] {$\potential$};
\end{scope}
\end{tikzpicture}
\end{center}
\caption{Given two players with preferences~$\preference_1,\preference_2$, one can construct the relation~$\joinofprefs$ and use it to construct the potential~$\potential$.
Here,~$\joinofprefs$ satisfies \cref{cond:jointly}, and satisfies \cref{cond:join_all} if uncomparable preferences are jointly communal.}
\label{fig:potential_example}
\end{figure}
\begin{small}
\begin{figure}[t]
\begin{subfigure}[b]{0.5\linewidth}
\begin{center}
    \begin{tikzpicture}
\node[blockfill](first) at (-2,0) {\begin{tikzcd}[row sep=small, column sep=tiny]
\mincltext
\end{tikzcd}};
\node[blockfill, right=0.1cm of first] (second) {
\begin{tikzcd}[row sep=small, column sep=tiny]
-\mincltext
\end{tikzcd}
};

\node[above=-0.1cm of first.north] {$\preference_1$};
\node[above=-0.1cm of second.north] {$\preference_2$};
\end{tikzpicture}
    \caption{Preferences for first case.}
    \label{fig:counterex_2_a}
\end{center}      
\end{subfigure}  
\begin{subfigure}[b]{0.5\linewidth}
\begin{center}
    \begin{tikzpicture}
\begin{scope}[scale=0.9, transform shape]
\node[blockfill](first) at (-2,0) {\begin{tikzcd}[row sep=small, column sep=tiny]
\jointmetric_1^1\\[-3pt]
\jointmetric_1^2\arrow[u,dash,solid,thick]
\end{tikzcd}};
\node[blockfill, right=0.1cm of first] (second) {
\begin{tikzcd}[row sep=small, column sep=tiny]
\jointmetric_2^2
\\[-3pt]
\jointmetric_2^1\arrow[u,dash,solid,thick]
\end{tikzcd}
};
\node[above=-0.1cm of first.north] {$\preference_1$};
\node[above=-0.1cm of second.north] {$\preference_2$};
\end{scope}
\end{tikzpicture}
    \caption{Preferences for second case.}
    \label{fig:counterex_1_a}
\end{center}      
\end{subfigure} \\[+5pt]
\begin{subfigure}[b]{0.4\linewidth}
\begin{center}
\begin{small}
    \begin{tabular}{c|cc}
    &$\traj_2^1$&$\traj_2^2$\\
    \hline
    $\traj_1^1$&$\tup{1,-1}$&$\tup{-1,1}$\\
    $\traj_1^2$&$\tup{-1,1}$&$\tup{1,-1}$
    \end{tabular}   
    \caption{Payoffs for first case.}
    \label{fig:counterex_2_b}
    \end{small}
\end{center}      
\end{subfigure}  
\begin{subfigure}[b]{0.55\linewidth}
\begin{small}
\begin{center}
        \setlength{\tabcolsep}{3pt}
        \begin{tabular}{c|cc}
        & \multicolumn{1}{c}{$\traj_2^1$} & \multicolumn{1}{c}{$\traj_2^2$} \\
        \hline
        $\traj_1^1$ &$\tup{ 
        \begin{tikzcd}[row sep=tiny]
        1\\
        1\arrow[u,dash,thick]
        \end{tikzcd},
        \begin{tikzcd}[row sep=tiny]
        1\\
        1\arrow[u,dash,thick]
        \end{tikzcd}}$ &  $\tup{
        \begin{tikzcd}[row sep=tiny]
        0\\
        0\arrow[u,dash,thick]
        \end{tikzcd},
        \begin{tikzcd}[row sep=tiny]
        1.5\\
        1\arrow[u,dash,thick]
        \end{tikzcd}
        }$\\
        $\traj_1^2$ &  $\tup{
        \begin{tikzcd}[row sep=tiny]
        0\\
        0\arrow[u,dash,thick]
        \end{tikzcd},
        \begin{tikzcd}[row sep=tiny]
        1.5\\
        1\arrow[u,dash,thick]
        \end{tikzcd}
        }$  & $\tup{
        \begin{tikzcd}[row sep=tiny]
        1\\
        1\arrow[u,dash,thick]
        \end{tikzcd},
        \begin{tikzcd}[row sep=tiny]
        1\\
        1\arrow[u,dash,thick]
        \end{tikzcd}
        }$
        \end{tabular}
    \caption{Payoffs for second case.}
    \label{fig:counterex_1_b}
\end{center}  
  \end{small}
\end{subfigure} 
\caption{
\emph{Violating \cref{cond:join_all}:} Consider the two player game defined in~\cref{fig:counterex_2_a,fig:counterex_2_b} .
Intuitively, if they choose the same action they are close together, different actions set them apart. Note that these preferences do not violate~\cref{cond:jointly}, but clearly violate~\cref{cond:join_all}.
A payoff matrix can be constructed as in \cref{fig:counterex_2_b}, which does not have a pure strategy \gls{abk:ne} (the game is equivalent to ``matching pennies'').
Violating \emph{\cref{cond:jointly}:} Consider a game with preferences depicted in \cref{fig:counterex_1_a}, which clearly violates \cref{cond:jointly}.
Let's assume that the preferences satisfy~\cref{cond:join_all}.
In this case, one can create a game which satisfies~\cref{cond:join_all}, but for which no pure strategy \gls{abk:ne} exists (\cref{fig:counterex_1_b}).
Technically, the preferences as given in \cref{fig:counterex_1_a} prevent~$\joinofprefs$ from being a poset, and hence prevent the construction of the potential as described in \cref{thm:existence_ne}.}
\label{fig:counterex_all}
\end{figure}
\end{small}
We further motivate~\cref{cond:join_all,cond:jointly} providing in~ \cref{fig:counterex_all} two counterexamples. Each violates only one condition, yet we can construct games where a pure equilibrium does not exist.

\subsection{Refining Equilibria via Preferences}\label{sec:refinement}

In this section we leverage the notion of refinement, introduced in \cref{sec:preliminaries}, and show how preference refinements allow one to refine \gls{abk:ne} of \glspl{abk:gpop}.
\cref{lem:min_antitone} and \cref{thm:refining_nes} encapsulate this concept.
\begin{lemma}
\label{lem:min_antitone}
Consider~$\preference=\tup{\metricsSet,\posleq}$ and~$\preference'=\tup{\metricsSet',\posleq'}$, and the respective induced preorders on the set of outcomes~$\tup{\outcomeset,\precsim}$,~$\tup{\outcomeset,\precsim'}$.
It holds:
\begin{equation*}
    \preference\precsim_{\pre{\allmetrics}} \preference' \implies \Min\tup{\outcomeset,\precsim}\supseteq \Min\tup{\outcomeset,\precsim'}.
\end{equation*}
\end{lemma}

\ifthenelse{\boolean{proofs}}{
\begin{proof}[Proof\ifthenelse{\boolean{proofs}}{}{
of \cref{lem:min_antitone}}]
Let~$p$ be a minimal element of~$ \tup{\outcomeset,\precsim'}$, implying that there is no~$q\in \outcomeset$ such that~$q\prec_\outcomeset' p$. 
\noindent Per absurdum, let's assume that~$p$ is not a minimal element of~$\tup{\outcomeset,\precsim}$, meaning that there is a~$q\in \outcomeset$ such that~$q\prec_\outcomeset p$. 
However, we know from \cref{def:preorder_preferences} that~$\tup{\outcomeset,\prec}\subseteq \tup{\outcomeset,\prec'}$, which implies that~$\tup{q,p}\in \tup{\outcomeset,\prec'}$.
This contradicts the starting assumption.
\end{proof}}{}
What is more, via \cref{lem:preorder_decisionspace} one can extend the result of \cref{lem:min_antitone} to minima of~$\tup{\trajset,\precsim}$.

\begin{theorem}
\label{thm:refining_nes}
Consider two \glspl{abk:gpop}~$\game_1,\game_2$ sharing the same constituents, exception made for players' preferences. We denote the preference of Player~$i$ in~$\game_j$ by~$\preference_i^j$.
It holds:
\begin{equation*}
    \begin{aligned}
    \preference_i^1\precsim_{\pre{\allmetrics}}\preference_i^2\quad \forall i\in \players \implies \neqweak(\game_1)\supseteq \neqweak(\game_2).
    \end{aligned}
\end{equation*}
\end{theorem}
In other words, \cref{thm:refining_nes} states that refining players' preferences in the sense of \cref{def:refinement} filters the \gls{abk:ne} of the game.
\ifthenelse{\boolean{proofs}}{
\begin{proof}[Proof\ifthenelse{\boolean{proofs}}{}{
of \cref{thm:refining_nes}}]
From \cref{lem:min_antitone} we know that~$\preference_i\precsim \preference'_i \quad \forall i\in \players \implies \Min\tup{\trajset,\precsim}\supseteq \Min\tup{\trajset,\precsim'}$. 
Therefore:
\begin{equation*}
\begin{aligned}
    \neqweak(\game)&=\{\traj \in \trajset \mid \traj_i\in \mathsf{BR}^{\precsim}_i(\traj_{-i}) \ \forall i\in \players\}\\
    &=\{\traj \in \trajset \mid \traj_i\in \Min\tup{\trajset_i,\precsim_{\trajset_i}} \ \forall i\in \players\}\\
    &\supseteq \{\traj \in \trajset \mid \traj_i\in \Min\tup{\trajset_i,\precsim_{\trajset_i}'} \ \forall i\in \players\}
    =\neqweak(\game').
\end{aligned}    
\end{equation*}
\end{proof}
\vspace{-0.5cm}}{}

\section{Trajectory Driving Game with Posetal Preferences}
\label{sec:first_case_study}
\begin{figure}[t]
   \centering
\begin{subfigure}[b]{\linewidth}
\begin{center}
    \input{tikz/pos_dg}
\end{center}    
    \caption{Scenario, trajectories, and equilibria of the case study.}
\end{subfigure}
~
\begin{subfigure}[b]{0.32\linewidth}
\begin{center}
\begin{tikzpicture}
\node[block, rounded corners=2pt, inner sep=0] at (0,0){
\includegraphics[width=\linewidth,angle=-90]{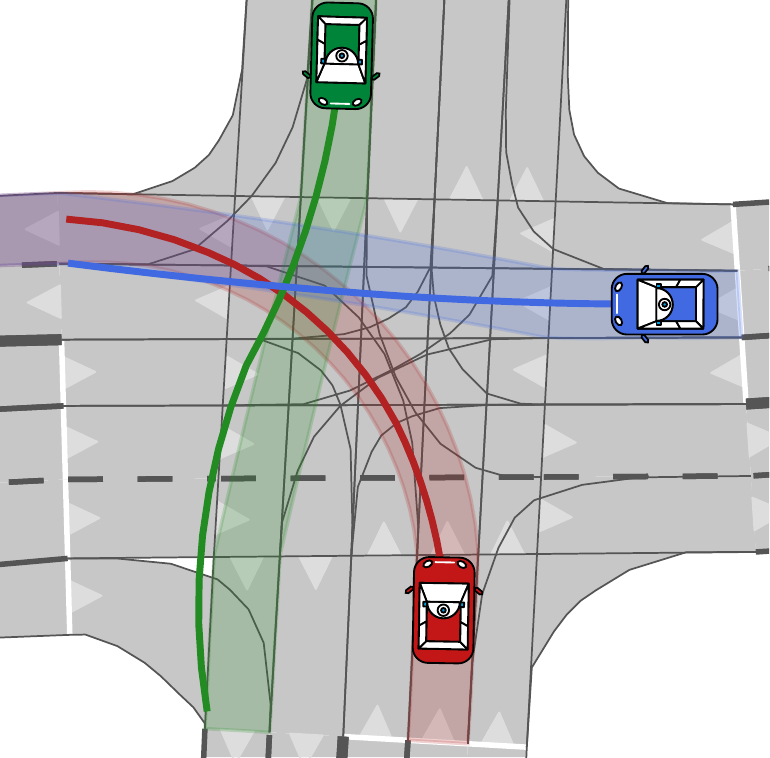}};
\end{tikzpicture} 
\end{center}
\caption{Strong \gls{abk:ne} \protect \circled{1}.}
\label{fig:strongne}
\end{subfigure}
\begin{subfigure}[b]{0.32\linewidth}
\begin{center}
\begin{tikzpicture}
\node[block,rounded corners=2pt, inner sep=0] at (0,0){
\includegraphics[width=\linewidth,angle=-90]{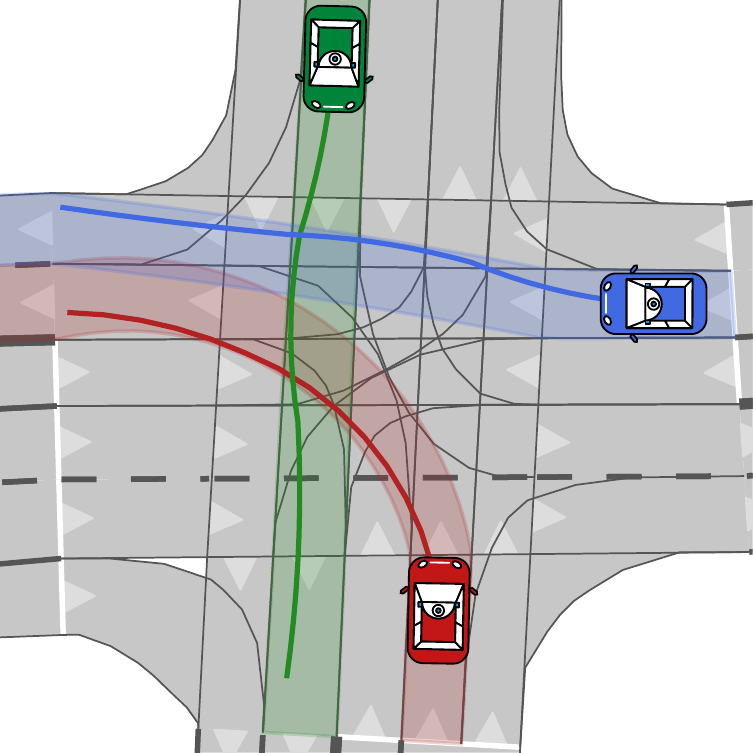}};
\end{tikzpicture}
\end{center}
\caption{Weak admissible \gls{abk:ne}~\protect\circled{2}.}
\label{fig:weakadmne}
\end{subfigure}
\begin{subfigure}[b]{0.32\linewidth}
\begin{center}
\begin{tikzpicture}
\node[block, rounded corners=2pt,inner sep=0] at (0,0){
\includegraphics[width=\linewidth,angle=-90]{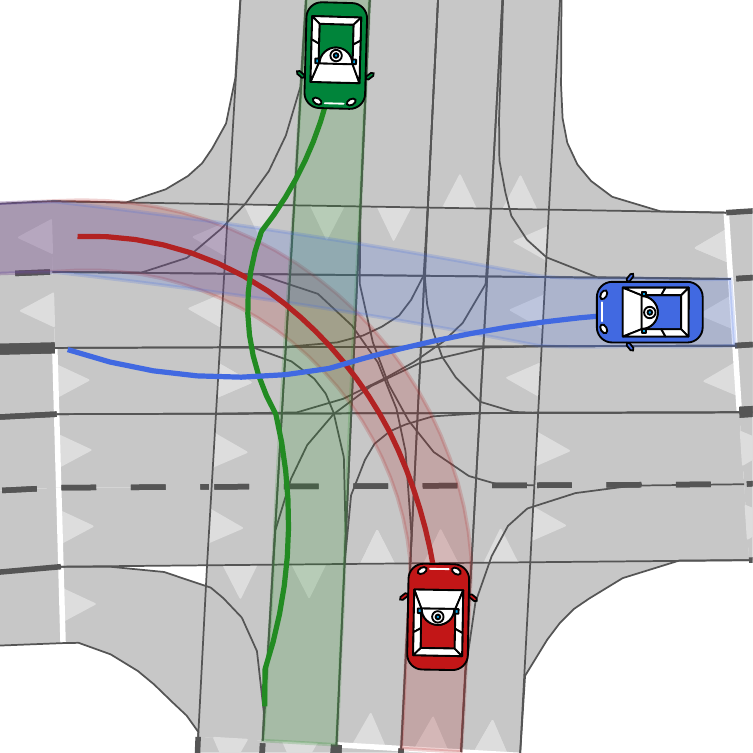}};
\end{tikzpicture}
\end{center}
\caption{Weak \gls{abk:ne} \protect \circled{3}.}
\label{fig:weakne}
\end{subfigure}

\caption{We report the benchmark scenario used to exemplify the introduced methodology (\textsc{USA\_Peach-4\_3\_T-1}) from CommonRoad~\cite{Althoff2017b}).
    The scenario is four-way intersection where each \gls{abk:av} chooses from a finite set of generated trajectories according to its preference.
    We generate around 90 trajectories for each \gls{abk:av}, as described in Paragraph a).
    Preferences are reported next to vehicles, and can be found in \cref{subfig:prefsne}.
    Specifically, we consider the same preference for Player~$\playertwo$ and~$\playerthree$, and a different one for~$\playerone$.
    \reviewed{The complete solution of the game is depicted in \cref{fig:equilibriarefinement}}.
    Three exemplary equilibria are reported in b), c), and d).}
    \label{fig:pos_dg}
\end{figure}

In this section, we evaluate the properties of~\glspl{abk:gpop} in an urban driving context. 
The players are \glspl{abk:av} solving the motion planning problem by choosing from a finite set of generated trajectories. 
The proposed setup is integrated on reproducible, publicly available CommonRoad scenarios~\cite{Althoff2017b}.

Given a scenario, we formulate a discrete trajectory game~\cite{zanardi2021udgs} as follows.
Each player comes with initial states and a pre-specified task to accomplish (e.g., navigate the intersection exiting WEST).
The corresponding action set consists of many trajectories, which are generated as described in Par. a).
The choice of the trajectory for the \gls{abk:av} is guided by the preference expressed as a poset of cost functions commonly used in motion planning. 
These include collision energy in case of an impact, drivable area violation, minimum clearance, time needed to reach the goal, progress along the prescribed reference, longitudinal and lateral comfort (function of the acceleration), lateral and heading deviation from the reference path.

\paragraph{Trajectory generation}
\label{par:trajgeneration}
In light of minimum-violation planning principles~\cite{Wongpiromsarn2020} we do not constrain the generation of trajectories to be \emph{already} compliant with the constraints and objectives of the players, for instance by considering only trajectories strictly compliant with the rules of the road.
We consider a single-track model for the vehicles and generate kinematically feasible trajectories around one or multiple reference paths assigned as goal directions. 
We promote heterogeneity of inputs and dispersion of states to not rule out a priori physically possible alternatives. 
For the sake of the example, we consider only trajectories which terminate the game, i.e., which reach the goal region. 
This is not a limitation of the method, which could compare arbitrary trajectories, but rather a simplification to compare \gls{abk:ne} of the \emph{same} game.
Promoting heterogeneity of trajectories is key for a minimum-violation planning approach to not miss out on possible solutions. 
We envision similar heuristics to be devised for game theoretic settings as well; yet, this is out of scope for this work. 
\subsection{Equilibria of the game}
Given a scenario, vehicles, tasks, and preferences, we compute the equilibria of the game performing an exhaustive search over the space of joint actions (trajectories) for different refinements of preferences.
As per~\cref{def:nasheq}, equilibria can either be strong or weak, and a subset of them could be admissible. 
In the following, we characterize and analyze equilibria for the setting reported in~\cref{fig:pos_dg}.
Specifically, the setting includes three players, i.e.,~$\agents=\{\playerone, \playertwo,\playerthree\}$ with respective goals and possible trajectories. The preferences are matched with~\cref{subfig:prefsne}.
Intuitively, the preferences of~$\playerone, \playertwo,\playerthree$ can be interpreted as follows.
All players prioritize the minimization of collision energy, but while~$\playertwo$ and~$\playerthree$ care more about preventing drivable area violation than preserving clearance and quickly progressing along a prescribed path,~$\playerone$ cares equally about drivable area violation and clearance, and prioritizes both over time.

We solve the game, and distinguish between strong and weak \gls{abk:ne}.
This game possesses 17 pure strategy weak \gls{abk:ne} and 1 pure strategy strong \gls{abk:ne} (\cref{subfig:9ne}).
Note that while the number of \gls{abk:ne} might seem large, we will describe formal methodologies to refine equilibria which have practical intuitions in~\cref{sec:refinement}.
Out of all the computed \gls{abk:ne}, only 10 are admissible.

Some comments are in order.
We graphically report three exemplary \gls{abk:ne}: a strong one (\circled{1}, \cref{fig:strongne}), a weak admissible one (\circled{2}, \cref{fig:weakadmne}), and a weak one (\circled{3}, \cref{fig:weakne}).

Payoffs for each player for selected equilibria are reported in \cref{fig:payoffs_a}.
In particular, we can see practical examples of equilibria dominating others.
For instance, consider equilibria \circled{2} (weak, admissible) and \circled{3} (weak, non-admissible). 
Clearly, \circled{2} dominates \circled{3} in any way: indeed, all agents are better off in drivable area violation (\cref{fig:payoffs_a}).

Finally, we are able to see \cref{lem:admissible_rank} at work.
By comparing equilibrium \circled{2} with equilibrium \circled{3}, we notice~$\rankk_i(\circled{3})\leq \rankk_i(\circled{2})$ for all~$i\in \agents$ (indeed,~$\rankk_i(\circled{3})=2$ and~$\rankk_i(\circled{2})=3$).

\begin{figure}[t]
    \begin{center}
    \begin{tikzpicture}
\begin{scope}[scale=0.7, transform shape]
\node[blockfill] (first) at (0,0){
     \begin{tikzcd}[column sep=tiny, row sep=small]
     &[-20pt]0&[-20pt]\\[-2pt]
     0\arrow[ur,thick,dash]&&0\arrow[ul,thick,dash]\\[-2pt]
     &8.8\arrow[ul, thick,dash]\arrow[ur, thick,dash]&\\[-2pt]
     1.3\arrow[ur, thick,dash]&&2.6\arrow[ul, thick,dash]\\[-2pt]
     &0.6\arrow[ul, thick, dash]\arrow[ur, thick, dash]&
     \end{tikzcd}};
     \node[blockfill, right=0.05cm of first] (second){ 
     \begin{tikzcd}[column sep=tiny, row sep=small]
     &[-20pt]0&[-20pt]\\[-2pt]
     &1.8\arrow[u, thick, dash]&\\[-2pt]
     1.1\arrow[ur,thick,dash]&&2.4\arrow[ul, thick,dash]\\[-2pt]
     &1.5\arrow[ul,thick,dash]\arrow[ur,thick,dash]&\\[-2pt]
     0.9\arrow[ur,thick,dash]&&1.8\arrow[ul,thick,dash]
     \end{tikzcd}};
     \node[blockfill, right=0.05cm of second] (third){ 
     \begin{tikzcd}[column sep=tiny, row sep=small]
     &[-20pt]0&[-20pt]\\[-2pt]
     &0\arrow[u, thick, dash]&\\[-2pt]
     1\arrow[ur,thick,dash]&&2.4\arrow[ul, thick,dash]\\[-2pt]
     &0.7\arrow[ul,thick,dash]\arrow[ur,thick,dash]&\\[-2pt]
    0.9\arrow[ur,thick,dash]&&0.2\arrow[ul,thick,dash]
     \end{tikzcd}};
     
\node[blockfill, right=0.1cm of third] (fourth) {
     \begin{tikzcd}[column sep=tiny, row sep=small]
     &[-20pt]0&[-20pt]\\[-2pt]
     0\arrow[ur,thick,dash]&&0\arrow[ul,thick,dash]\\[-2pt]
     &5.6\arrow[ul, thick,dash]\arrow[ur, thick,dash]&\\[-2pt]
     0.4\arrow[ur, thick,dash]&&2.6\arrow[ul, thick,dash]\\[-2pt]
     &0.8\arrow[ul, thick, dash]\arrow[ur, thick, dash]&
     \end{tikzcd}};
     \node[blockfill, right=0.05cm of fourth] (fifth){ 
     \begin{tikzcd}[column sep=tiny, row sep=small]
     &[-20pt]0&[-20pt]\\[-2pt]
     &5.4\arrow[u, thick, dash]&\\[-2pt]
     0\arrow[ur,thick,dash]&&0\arrow[ul, thick,dash]\\[-2pt]
     &1.2\arrow[ul,thick,dash]\arrow[ur,thick,dash]&\\[-2pt]
     0.9\arrow[ur,thick,dash]&&1.4\arrow[ul,thick,dash]
     \end{tikzcd}};
     \node[blockfill, right=0.05cm of fifth] (sixth){ 
     \begin{tikzcd}[column sep=tiny, row sep=small]
     &[-20pt]0&[-20pt]\\[-2pt]
     &0\arrow[u, thick, dash]&\\[-2pt]
     5.2\arrow[ur,thick,dash]&&0\arrow[ul, thick,dash]\\[-2pt]
     &1.6\arrow[ul,thick,dash]\arrow[ur,thick,dash]&\\[-2pt]
    0.9\arrow[ur,thick,dash]&&1.2\arrow[ul,thick,dash]
     \end{tikzcd}}; 
     
\node[blockfill, right=0.1cm of sixth] (seventh) {
     \begin{tikzcd}[column sep=tiny, row sep=small]
     &[-20pt]0&[-20pt]\\[-2pt]
     7.7\arrow[ur,thick,dash]&&0\arrow[ul,thick,dash]\\[-2pt]
     &10\arrow[ul, thick,dash]\arrow[ur, thick,dash]&\\[-2pt]
     0.9\arrow[ur, thick,dash]&&2.7\arrow[ul, thick,dash]\\[-2pt]
     &0.6\arrow[ul, thick, dash]\arrow[ur, thick, dash]&
     \end{tikzcd}};
     \node[blockfill, right=0.05cm of seventh] (eigth){ 
     \begin{tikzcd}[column sep=tiny, row sep=small]
     &[-20pt]0&[-20pt]\\[-2pt]
     &5.9\arrow[u, thick, dash]&\\[-2pt]
     5.8\arrow[ur,thick,dash]&&6\arrow[ul, thick,dash]\\[-2pt]
     &3\arrow[ul,thick,dash]\arrow[ur,thick,dash]&\\[-2pt]
     1.3\arrow[ur,thick,dash]&&3.7\arrow[ul,thick,dash]
     \end{tikzcd}};
     \node[blockfill, right=0.05cm of eigth] (nineth){ 
     \begin{tikzcd}[column sep=tiny, row sep=small]
     &[-20pt]0&[-20pt]\\[-2pt]
     &10.4\arrow[u, thick, dash]&\\[-2pt]
     9.09\arrow[ur,thick,dash]&&1.1\arrow[ul, thick,dash]\\[-2pt]
     &2.6\arrow[ul,thick,dash]\arrow[ur,thick,dash]&\\[-2pt]
    0.9\arrow[ur,thick,dash]&&1.4\arrow[ul,thick,dash]
     \end{tikzcd}};   
\end{scope}  
\draw [decorate,decoration={brace,amplitude=6pt},yshift=0pt, thick] (third.south east) -- (first.south west) node [black,midway,xshift=-0.6cm] {};
\draw [decorate,decoration={brace,amplitude=6pt},yshift=0pt, thick] (sixth.south east) -- (fourth.south west) node [black,midway,xshift=-0.6cm] {};
\draw [decorate,decoration={brace,amplitude=6pt},yshift=0pt, thick] (nineth.south east) -- (seventh.south west) node [black,midway,xshift=-0.6cm] {};
     \node[below=0.1cm of second.south] {equilibrium \circled{1}};
     \node[below=0.1cm of fifth.south] {equilibrium \circled{2}};
     \node[below=0.1cm of eigth.south] {equilibrium \circled{3}};
     \node[above=-0.1cm of first.north] {$\playerone$};
     \node[above=-0.1cm of second.north] {$\playertwo$};
     \node[above=-0.1cm of third.north] {$\playerthree$};
     \node[above=-0.1cm of fourth.north] {$\playerone$};
     \node[above=-0.1cm of fifth.north] {$\playertwo$};
     \node[above=-0.1cm of sixth.north] {$\playerthree$};
     \node[above=-0.1cm of seventh.north] {$\playerone$};
     \node[above=-0.1cm of eigth.north] {$\playertwo$};
     \node[above=-0.1cm of nineth.north] {$\playerthree$};
\end{tikzpicture}
    \caption{Selected payoffs for equilibria of the game depicted in \cref{fig:pos_dg}. 
    \gls{abk:ne} \protect \circled{1} is strong, \protect \circled{2} is weak and admissible, and \protect \circled{3} is weak and non-admissible.}
    \label{fig:payoffs_a}
    \end{center}
    \vspace{-1cm}
\end{figure}

\subsection{Refinement in a driving game with posetal preferences}

\begin{figure*}[t]
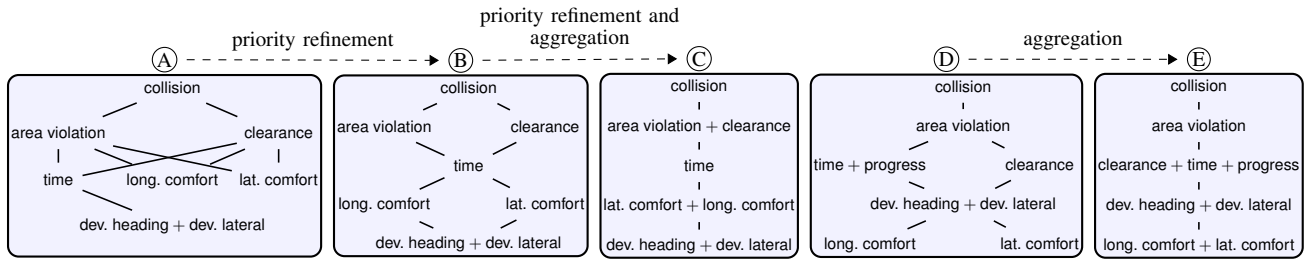
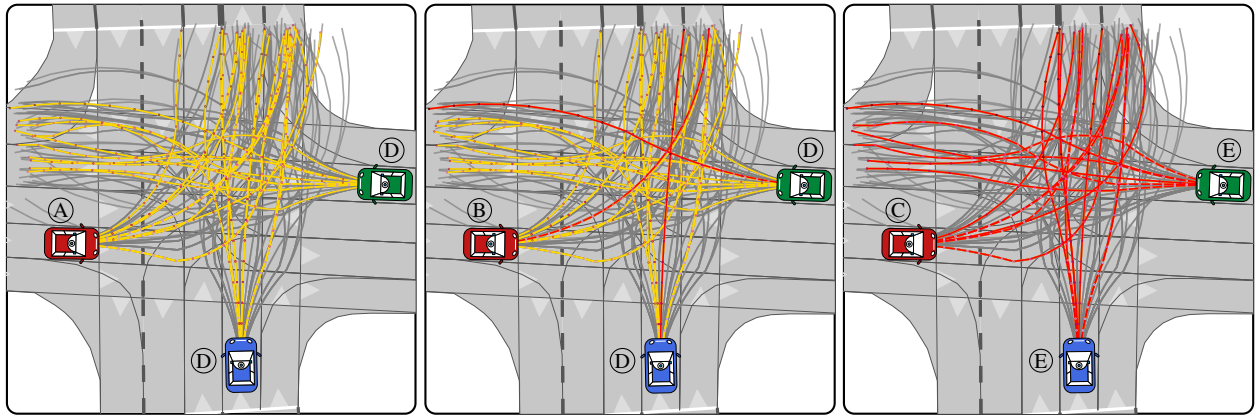

    \centering
    \begin{subfigure}[b]{\textwidth}
    \centering
    \begin{tikzpicture}
     \begin{scope}[scale =0.75, transform shape]
     \node[blockfill] (first) at (0,0){
     \begin{tikzcd}[column sep=tiny, row sep=small]
     &[-25pt]\colltext&[-25pt]\\
     \drivtext\arrow[ur, thick, dash]&&\mincltext\arrow[ul,thick,dash]\\
     \epitext\arrow[u, thick, dash]\arrow[urr, thick, dash]&\longitext\arrow[ur, thick, dash]\arrow[ul, thick, dash]&\lattext\arrow[ull, thick, dash]\arrow[u, thick, dash]\\
     &\devhtext+\devltext\arrow[ul,dash,thick]&\\[+2pt]
     \end{tikzcd}};
     
     \node[blockfill, right=0.2cm of first] (third) {
     \begin{tikzcd}[column sep=tiny, row sep=tiny]
     &[-40pt]\colltext&[-40pt]\\
     \drivtext\arrow[ur, thick, dash]&&\mincltext\arrow[ul,thick,dash]\\
     &\epitext\arrow[ul, thick, dash]\arrow[ur, thick, dash]&\\
     \longitext\arrow[ur, thick, dash]&&\lattext\arrow[ul, thick, dash]\\
     &\devhtext+\devltext\arrow[ul,dash,thick]\arrow[ur,dash,thick]&
     \end{tikzcd}};
          \node[blockfill, right=0.2cm of third] (fourth) {
     \begin{tikzcd}[column sep=tiny, row sep=small]
     \colltext\\[-3pt]
     \drivtext+\mincltext\arrow[u, thick, dash]\\[-3pt]
     \epitext \arrow[u, thick, dash]\\[-3pt]
     \lattext+\longitext \arrow[u,thick,dash]\\[-3pt]
     \devhtext+\devltext \arrow[u,thick,dash]
     \end{tikzcd}};
     \node[blockfill, right=0.2cm of fourth] (second){
     \begin{tikzcd}[column sep=tiny, row sep=tiny]
     &[-40pt]\colltext&[-40pt]\\
     &\drivtext\arrow[u, thick, dash]&\\
     \epitext+\progtext\arrow[ur,thick,dash]&&\mincltext\arrow[ul, thick,dash]\\
     &\devhtext+\devltext\arrow[ul,thick,dash]\arrow[ur,thick,dash]&\\
     \longitext\arrow[ur,thick,dash]&&\lattext\arrow[ul,thick,dash]
     \end{tikzcd}};
     \node[blockfill, right=0.2cm of second] (fifth) {
     \begin{tikzcd}[column sep=small, row sep=small]
     \colltext\\[-3pt]
     \drivtext\arrow[u,thick,dash]\\[-3pt]
     \mincltext+\epitext+\progtext\arrow[u, thick, dash]\\[-3pt]
     \devhtext+\devltext\arrow[u, thick, dash]\\[-3pt]
     \longitext+\lattext\arrow[u, thick, dash]
     \end{tikzcd}};
     \end{scope}
     \node[above=-0.1cm of first] (A) {$\circled{A}$}; 
     \node[above=-0.1cm of second] (B) {$\circled{D}$}; 
     \node[above=-0.1cm of third] (C) {$\circled{B}$}; 
     \node[above=-0.1cm of fourth] (D) {$\circled{C}$}; 
     \node[above=-0.1cm of fifth] (E) {$\circled{E}$}; 
     \draw[-Triangle, dashed] (A)-- node[pos=0.5,above] {priority refinement} (C);
     \draw[-Triangle, dashed] (C)-- node[pos=0.5,above] {\begin{tabular}{c}priority refinement and\\ aggregation\end{tabular}} (D);
     \draw[-Triangle, dashed] (B)-- node[pos=0.5,above] {\text{aggregation}} (E);

     \end{tikzpicture}
    \caption{Preferences used in the case study and their refinement via \cref{def:priority_ref_op,def:aggregation_operation}.}
    \label{subfig:prefsne}
    \end{subfigure}\\[+5pt]
    \begin{subfigure}[b]{0.3\textwidth}
    \input{tikz/ref_bis_1.tikz}
    \vspace{-0.35cm}
    \caption{Preference in~\cref{subfig:prefsne} yields 28 weak \gls{abk:ne}, no strong \gls{abk:ne}, and 17 admissible \gls{abk:ne}.}
     \label{subfig:60ne}
    \end{subfigure}
    \begin{subfigure}[b]{0.3\textwidth}
    \input{tikz/ref_bis_2.tikz}
    \vspace{-0.35cm}
    \caption{Preference in~\cref{subfig:prefsne} yields 16 weak \gls{abk:ne}, 1 strong \gls{abk:ne}, and 10 admissible \gls{abk:ne}.}
     \label{subfig:9ne}
    \end{subfigure}
    \begin{subfigure}[b]{0.3\textwidth}
    \input{tikz/ref_bis_3.tikz}
    \vspace{-0.35cm}
    \caption{Preference in~\cref{subfig:prefsne} yields no weak \gls{abk:ne}, 7 strong \gls{abk:ne}, and 3 admissible \gls{abk:ne}.}
     \label{subfig:6ne}
    \end{subfigure}
     \caption{From left to right, we observe the \emph{same} game played with \emph{refined} preferences.
     Strong \gls{abk:ne} are depicted in red, and weak \gls{abk:ne} in yellow.
     From (a) to (b)~$\playerone$ refines her preferences via \cref{def:priority_ref_op}, and from (b) to (c)~$\playerone$ refines her preferences via \cref{def:priority_ref_op} and \cref{def:aggregation_operation}, and~$\playertwo, \playerthree$ refine their preferences via \cref{def:priority_ref_op} and \cref{def:aggregation_operation}.
     Further refining the preference shrinks the set of \gls{abk:ne} of the game.}
     \label{fig:equilibriarefinement}
     \vspace{-0.3cm}
\end{figure*}

We now showcase the importance of \cref{thm:refining_nes}, instantiating it in the case study reported in \cref{sec:first_case_study}.
We report all the preferences used within the case study, as well as the refinement operations relating them in \cref{subfig:prefsne}.
We start with the setting depicted in \cref{subfig:60ne}, in which players~$\playertwo$ and~$\playerthree$ share the same preference structure \circled{D}. 
The specific game possesses 28 weak \gls{abk:ne}, no strong \gls{abk:ne}, and 17 admissible \gls{abk:ne}. 
We depict strong \gls{abk:ne} in red and weak \gls{abk:ne} in yellow.
From here, we refine the preference of~$\playerone$ from \circled{A} to \circled{B}, by repeatedly applying a priority refinement operation (\cref{def:priority_ref_op}). 
The resulting game possesses 16 weak \gls{abk:ne}, 1 strong \gls{abk:ne}, and 10 admissible \gls{abk:ne}, highlighting the validity of \cref{thm:refining_nes} (indeed, by refining~$\playerone$'s preference we are shrinking the set of weak \gls{abk:ne} of the game).
We further refine the preferences, now both for~$\playerone$ and~$\playertwo,\playerthree$.
For~$\playerone$, we aggregate area violation with clearance and longitudinal comfort with lateral comfort via \cref{def:aggregation_operation}, and get \circled{C}. 
Similarly, for~$\playertwo,\playerthree$, we aggregate clearance with time and progress along a given reference, and longitudinal with lateral comfort, obtaining \circled{E}.
The resulting game possesses no weak \gls{abk:ne}, 7 strong \gls{abk:ne}, and 3 admissible \gls{abk:ne}, again agreeing with \cref{thm:refining_nes}.

\subsection{Discussion}
The presented case study has pedagogical value and practically showcases all the theoretical results. 
Notably, we show how the admissible \gls{abk:ne} represent the ``fairest'' minimum violation for the agents' preferences and how the preference refinement shrinks the set of solutions. 
The richness of posetal preferences allows one to rethink standard game-theoretical tools. 
For instance, the concept of security strategies can be considered in relation to an arbitrary rank of the others' preferences. 
That is, a player could assume that the others play rational only up to rank~$k$. 

The presented case has practical limitations due to the combinatorial explosion of joint actions. 
We envision two ways to address this.
The first is to consider a game played in a receding horizon fashion, over a shorter time horizon; this will curb significantly the action space. 
The second option is to solve a scalarized version of the game (e.g., enumerating the homotopy classes of who goes first) with different techniques and then check how the solutions are ranked with respect to the original preferences.
In this regard, we highlight that the posetal preferences provide a formidable checker for the ``quality'' of equilibria, which could be provided by different systems: an oracle, approximate solution methods, or learning methods.
\section{Conclusions}
In this work, we presented games in which players express a preference on outcomes via a prioritized hierarchy of metrics. 
This aims at modeling the interactive aspect of multi-robot settings in combination with the multi-objective nature of mobile robots. 
Particularly, for modern applications which often requires non-trivial behavior specification.

We extended classic game theoretic constructions to this kind of games, and provided sufficient conditions for the existence of pure \gls{abk:ne} for finite actions. 
Furthermore, we demonstrated that one can systematically refine the set of equilibria performing interpretable operations on the preferences (e.g., aggregating two metrics).
Finally, we instantiated the developed methodologies in the practical case of trajectory selection for \glspl{abk:av} in urban driving scenarios.

This work opens various avenues for future research.
First, posetal preferences offer a formidable tool that generalizes relevant aspects of decision making problems for robots.
For instance, the proposed framework could include objectives specified in linear temporal logic~\cite{Tumova2013}.
\reviewed{Second, the framework could be extended to include a bit of ``slackness'' in the preference structures, e.g., considering \emph{lexicographic semiorders}~\cite{Manzini2012a}. The addition of a indifference threshold prevents a small noise in high priority metrics to be detrimental for lower metrics. For example, if lane keeping is more important than comfort, still one can allow to trade off a small lane violation if it helps avoiding a hard braking.}

Third, while this work develops a thorough understanding of posetal games, it poses new challenges on the computational side of things. 
Particularly, this entails the extension of common optimization tools (e.g., optimization on continuous decision spaces) to the case of preordered sets.
Finally, the monotonicity properties of posetal games in terms of equilibria refinement provide an ideal setting for the development of dynamic, estimation-based techniques. 
Each preference can be seen as a specific type of player, laying the foundations for Bayesian frameworks.
In particular, for the application of self-driving cars, we recognize that high rank objectives are common to all \emph{rational} players, while the specific preference of players can be learned and refined dynamically via the single interactions.

\bibliographystyle{IEEEtran}
\bibliography{references_mendeley}

\begin{thebibliography}{10}
\def\url#1{}
\csname url@rmstyle\endcsname
\providecommand{\newblock}{\relax}
\providecommand{\bibinfo}[2]{#2}
\providecommand\BIBentrySTDinterwordspacing{\spaceskip=0pt\relax}
\providecommand\BIBentryALTinterwordstretchfactor{4}
\providecommand\BIBentryALTinterwordspacing{\spaceskip=\fontdimen2\font plus
\BIBentryALTinterwordstretchfactor\fontdimen3\font minus
  \fontdimen4\font\relax}
\providecommand\BIBforeignlanguage[2]{{%
\expandafter\ifx\csname l@#1\endcsname\relax
\typeout{** WARNING: IEEEtran.bst: No hyphenation pattern has been}%
\typeout{** loaded for the language `#1'. Using the pattern for}%
\typeout{** the default language instead.}%
\else
\language=\csname l@#1\endcsname
\fi
#2}}

\bibitem{Hughes2017WhenWrong}
\BIBentryALTinterwordspacing
J.~Hughes and A.~A. Scholer, ``{When Wanting the Best Goes Right or Wrong},''
  \emph{Personality and Social Psychology Bulletin}, vol.~43, no.~4, pp.
  570--583, 4 2017.
  \url{http://journals.sagepub.com/doi/10.1177/0146167216689065}
\BIBentrySTDinterwordspacing

\bibitem{Censi2019b}
\BIBentryALTinterwordspacing
A.~Censi, K.~Slutsky, T.~Wongpiromsarn, D.~Yershov, S.~Pendleton, J.~Fu, and
  E.~Frazzoli, ``{Liability, Ethics, and Culture-Aware Behavior Specification
  using Rulebooks},'' in \emph{2019 International Conference on Robotics and
  Automation (ICRA)}.\hskip 1em plus 0.5em minus 0.4em\relax IEEE, 5 2019, pp.
  8536--8542.  \url{https://ieeexplore.ieee.org/document/8794364/
  http://arxiv.org/abs/1902.09355}
\BIBentrySTDinterwordspacing

\bibitem{Trautman2010}
P.~Trautman and A.~Krause, ``{Unfreezing the robot: Navigation in dense,
  interacting crowds},'' \emph{IEEE/RSJ 2010 International Conference on
  Intelligent Robots and Systems, IROS 2010 - Conference Proceedings}, pp.
  797--803, 2010.

\bibitem{schwarting2018planning}
\BIBentryALTinterwordspacing
W.~Schwarting, J.~Alonso-Mora, and D.~Rus, ``{Planning and Decision-Making for
  Autonomous Vehicles},'' \emph{Annual Review of Control, Robotics, and
  Autonomous Systems}, vol.~1, no.~1, pp. 187--210, 5 2018.
  \url{https://www.annualreviews.org/doi/10.1146/annurev-control-060117-105157}
\BIBentrySTDinterwordspacing

\bibitem{Sadigh2018}
\BIBentryALTinterwordspacing
D.~Sadigh, N.~Landolfi, S.~S. Sastry, S.~A. Seshia, and A.~D. Dragan,
  ``{Planning for cars that coordinate with people: leveraging effects on human
  actions for planning and active information gathering over human internal
  state},'' \emph{Autonomous Robots}, vol.~42, no.~7, 10 2018.
  \url{https://doi.org/10.1007/s10514-018-9746-1
  http://link.springer.com/10.1007/s10514-018-9746-1}
\BIBentrySTDinterwordspacing

\bibitem{Dreves2018}
\BIBentryALTinterwordspacing
A.~Dreves and M.~Gerdts, ``{A generalized Nash equilibrium approach for optimal
  control problems of autonomous cars},'' \emph{Optimal Control Applications
  and Methods}, vol.~39, no.~1, pp. 326--342, 1 2018.
  \url{http://doi.wiley.com/10.1002/oca.2348}
\BIBentrySTDinterwordspacing

\bibitem{Fisac2019}
\BIBentryALTinterwordspacing
J.~F. Fisac, E.~Bronstein, E.~Stefansson, D.~Sadigh, S.~S. Sastry, and A.~D.
  Dragan, ``{Hierarchical Game-Theoretic Planning for Autonomous Vehicles},''
  \emph{2019 International Conference on Robotics and Automation (ICRA)}, vol.
  2019-May, pp. 9590--9596, 10 2018.
  \url{https://ieeexplore.ieee.org/document/8794007/
  http://arxiv.org/abs/1810.05766}
\BIBentrySTDinterwordspacing

\bibitem{Fridovich-Keil2020}
\BIBentryALTinterwordspacing
D.~Fridovich-Keil, V.~Rubies-Royo, and C.~J. Tomlin, ``{An Iterative Quadratic
  Method for General-Sum Differential Games with Feedback Linearizable
  Dynamics},'' in \emph{2020 IEEE International Conference on Robotics and
  Automation (ICRA)}.\hskip 1em plus 0.5em minus 0.4em\relax IEEE, 5 2020, pp.
  2216--2222.  \url{https://ieeexplore.ieee.org/document/9196517/}
\BIBentrySTDinterwordspacing

\bibitem{Fridovich-Keil2020a}
D.~Fridovich-Keil, A.~Bajcsy, J.~F. Fisac, S.~L. Herbert, S.~Wang, A.~D.
  Dragan, and C.~J. Tomlin, ``{Confidence-aware motion prediction for real-time
  collision avoidance},'' \emph{International Journal of Robotics Research},
  vol.~39, no. 2-3, pp. 250--265, 2020.

\bibitem{Tian2020Game-TheoreticValidation}
R.~Tian, N.~Li, I.~Kolmanovsky, Y.~Yildiz, and A.~R. Girard, ``{Game-Theoretic
  Modeling of Traffic in Unsignalized Intersection Network for Autonomous
  Vehicle Control Verification and Validation},'' \emph{IEEE Transactions on
  Intelligent Transportation Systems}, pp. 1--16, 2020.

\bibitem{Schwarting2019a}
W.~Schwarting, A.~Pierson, J.~Alonso-Mora, S.~Karaman, and D.~Rus, ``{Social
  behavior for autonomous vehicles},'' \emph{Proceedings of the National
  Academy of Sciences of the United States of America}, vol. 116, no.~50, pp.
  2492--24\,978, 2019.

\bibitem{LeCleach2021}
\BIBentryALTinterwordspacing
S.~Le~Cleach, M.~Schwager, and Z.~Manchester, ``{LUCIDGames: Online Unscented
  Inverse Dynamic Games for Adaptive Trajectory Prediction and Planning},''
  \emph{IEEE Robotics and Automation Letters}, vol.~6, no.~3, pp. 5485--5492,
  2021.  \url{https://arxiv.org/abs/2011.08152}
\BIBentrySTDinterwordspacing

\bibitem{Brito2021}
\BIBentryALTinterwordspacing
B.~Brito, A.~Agarwal, and J.~Alonso-Mora, ``{Learning Interaction-aware
  Guidance Policies for Motion Planning in Dense Traffic Scenarios},'' 7 2021.
  \url{http://arxiv.org/abs/2107.04538}
\BIBentrySTDinterwordspacing

\bibitem{Ding2021EPSILON:Environments}
\BIBentryALTinterwordspacing
W.~Ding, L.~Zhang, J.~Chen, and S.~Shen, ``{EPSILON: An Efficient Planning
  System for Automated Vehicles in Highly Interactive Environments},''
  \emph{IEEE Transactions on Robotics}, pp. 1--21, 8 2021.
  \url{http://arxiv.org/abs/2108.07993
  https://ieeexplore.ieee.org/document/9526613/}
\BIBentrySTDinterwordspacing

\bibitem{Wongpiromsarn2020}
\BIBentryALTinterwordspacing
T.~Wongpiromsarn, K.~Slutsky, E.~Frazzoli, and U.~Topcu, ``{Minimum-Violation
  Planning for Autonomous Systems: Theoretical and Practical Considerations},''
  in \emph{2021 American Control Conference (ACC)}.\hskip 1em plus 0.5em minus
  0.4em\relax IEEE, 5 2021, pp. 4866--4872.
  \url{https://arxiv.org/abs/2009.11954 http://arxiv.org/abs/2009.11954
  https://ieeexplore.ieee.org/document/9483174/}
\BIBentrySTDinterwordspacing

\bibitem{Collin2020}
A.~Collin, A.~Bilka, S.~Pendleton, and R.~D. Tebbens, ``{Safety of the Intended
  Driving Behavior Using Rulebooks},'' \emph{IEEE Intelligent Vehicles
  Symposium, Proceedings}, no.~Iv, pp. 136--143, 2020.

\bibitem{zanardi2021udgs}
\BIBentryALTinterwordspacing
A.~Zanardi, E.~Mion, M.~Bruschetta, S.~Bolognani, A.~Censi, and E.~Frazzoli,
  ``{Urban Driving Games with Lexicographic Preferences and Socially Efficient
  Nash Equilibria},'' \emph{IEEE Robotics and Automation Letters}, pp. 1--1,
  2021.  \url{https://ieeexplore.ieee.org/document/9385938/}
\BIBentrySTDinterwordspacing

\bibitem{BassamHelou}
\BIBentryALTinterwordspacing
B.~Helou, A.~Dusi, A.~Collin, N.~Mehdipour, Z.~Chen, C.~Lizarazo, C.~Belta,
  T.~Wongpiromsarn, R.~D. Tebbens, and O.~Beijbom, ``{The Reasonable Crowd:
  Towards evidence-based and interpretable models of driving behavior},''
  no.~1, 7 2021.  \url{http://arxiv.org/abs/2107.13507}
\BIBentrySTDinterwordspacing

\bibitem{luce1956}
D.~Luce, ``{Semiorders and a Theory of Utility Discrimination},''
  \emph{Econometrica}, vol.~24, no.~2, pp. 178--191, 1956.

\bibitem{Shapley1974GameUtility}
L.~S. Shapley and M.~Shubik, ``{Game Theory in Economics, Preferences and
  Utility},'' 1974, ch.~4.

\bibitem{Rozen2018}
V.~V. Rozen, ``{Games with Ordered Outcomes},'' \emph{Journal of Mathematical
  Sciences (United States)}, vol. 235, no.~6, pp. 740--755, 2018.

\bibitem{LeRoux2008}
S.~Le~Roux, ``{Generalisation and Formalisation in Game Theory},'' Ph.D.
  dissertation, 2008.

\bibitem{Althoff2017b}
M.~Althoff, M.~Koschi, and S.~Manzinger, ``{CommonRoad: Composable benchmarks
  for motion planning on roads},'' \emph{IEEE Intelligent Vehicles Symposium,
  Proceedings}, pp. 719--726, 2017.

\bibitem{PriestleyIntroductionOrder}
H.~Priestley, \emph{{Introduction to Lattices and Order}}, 2nd~ed.

\bibitem{Nash1950}
J.~Nash, ``{Non-Cooperative Games},'' Ph.D. dissertation, Princeton Universtiy,
  1950.

\bibitem{Monderer1996}
\BIBentryALTinterwordspacing
D.~Monderer and L.~S. Shapley, ``{Potential Games},'' \emph{Games and Economic
  Behavior}, vol.~14, no.~1, pp. 124--143, 5 1996.
  \url{https://linkinghub.elsevier.com/retrieve/pii/S0899825696900445}
\BIBentrySTDinterwordspacing

\bibitem{Hespanha2017}
J.~P. Hespanha, \emph{{Noncooperative game theory : an introduction for
  engineers and computer scientists}}.\hskip 1em plus 0.5em minus 0.4em\relax
  Princeton University Press, 2017.

\bibitem{Tumova2013}
\BIBentryALTinterwordspacing
J.~Tumova, L.~I.~R. Castro, S.~Karaman, E.~Frazzoli, and D.~Rus,
  ``{Minimum-violation LTL planning with conflicting specifications},'' in
  \emph{2013 American Control Conference}.\hskip 1em plus 0.5em minus
  0.4em\relax IEEE, 6 2013, pp. 200--205.  \url{http://arxiv.org/abs/1303.3679
  http://ieeexplore.ieee.org/document/6579837/}
\BIBentrySTDinterwordspacing

\bibitem{Manzini2012a}
\BIBentryALTinterwordspacing
P.~Manzini and M.~Mariotti, ``{Choice by lexicographic semiorders},''
  \emph{Theoretical Economics}, vol.~7, no.~1, pp. 1--23, 2012.
  \url{http://doi.wiley.com/10.3982/TE679}
\BIBentrySTDinterwordspacing

\end{thebibliography}
\end{document}